\documentclass[aip,jcp,amsmath,amssymb,preprint]{revtex4-2}

\usepackage{graphicx}
\usepackage{dcolumn}
\usepackage{bm}
\usepackage[mathlines]{lineno}
\usepackage{tikz}
\usepackage{multirow}
\usepackage[utf8]{inputenc}
\usepackage[T1]{fontenc}
\usepackage{mathptmx}
\usepackage[hidelinks]{hyperref}
\usepackage{xcolor}
\usepackage{etoolbox}
\usepackage[caption=false]{subfig}

\makeatletter
\def\@email#1#2{%
 \endgroup
 \patchcmd{\titleblock@produce}
  {\frontmatter@RRAPformat}
  {\frontmatter@RRAPformat{\produce@RRAP{*#1\href{mailto:#2}{#2}}}\frontmatter@RRAPformat}
  {}{}
}%
\makeatother

\newcommand{\ket}[1]{\ensuremath{| #1 \rangle}}
\newcommand{\bra}[1]{\ensuremath{\langle #1 |}}
\newcommand{\braket}[2]{\ensuremath{\langle #1 | #2 \rangle}}
\newcommand{\ii}{\ensuremath{\text{i}}}
\newcommand{\ud}{\ensuremath{\text{d}}}
\newcommand{\vect}[1]{{\bf #1}}
\newcommand{\cm}{\ensuremath{\text{cm}^{-1}}}

\newcommand{\tr}{\ensuremath{\text{Tr}}}

\newcommand{\boltz}{\ensuremath{k_\text{B}}}

\begin{document}


\title{Modeling the emission spectra of polycyclic aromatic hydrocarbons by recurrent fluorescence}
\author{Damien Borja}
\affiliation{Université Paris-Saclay, CNRS, Institut des Sciences Moléculaires d'Orsay, 91405, Orsay, France}
\affiliation{Université Grenoble-Alpes, CNRS, LIPhy, 38000 Grenoble, France}
\author{Florent Calvo}
\affiliation{Université Grenoble-Alpes, CNRS, LIPhy, 38000 Grenoble, France}
\author{Pascal Parneix}
\affiliation{Université Paris-Saclay, CNRS, Institut des Sciences Moléculaires d'Orsay, 91405, Orsay, France}
\author{Cyril Falvo}
\email{cyril.falvo@universite-paris-saclay.fr}
\affiliation{Université Paris-Saclay, CNRS, Institut des Sciences Moléculaires d'Orsay, 91405, Orsay, France}
\affiliation{Université Grenoble-Alpes, CNRS, LIPhy, 38000 Grenoble, France}

\date{\today}

\begin{abstract}
Recurrent fluorescence (RF) is an important relaxation mechanism in polycyclic aromatic hydrocarbons (PAHs), which could stabilize them and contribute to the production of aromatic infrared bands that are observed in the infrared spectra of the interstellar medium (ISM). In this theoretical work, a statistical model of relaxation by recurrent fluorescence is formally developed, including Herzberg-Teller and Duschinsky rotation effects as well as a full account of vibrational progressions. Using canonical and harmonic approximations, the RF rate constants can be determined from the transition dipole moment time autocorrelation functions. Application to the naphthalene, anthracene, and pyrene cations is presented based on quantum chemical inputs obtained from time-dependent density-functional theory. For these highly symmetric molecules, the low-lying, symmetry-forbidden electronic transitions are predicted to contribute possibly even more than higher energy, non-forbidden transitions. Such an unexpected contribution could increase the cooling efficiency of PAHs and, in turn, stabilize them further under the highly ionized environments of the ISM.
\end{abstract}

\maketitle

\section{Introduction}

The presence of polycyclic aromatic hydrocarbons (PAHs) in the interstellar medium (ISM) was hypothesized over 40 years ago to explain the observation of the so-called aromatic infrared bands (AIBs).\cite{leger1984identification,Allamandola:1985uq} This hypothesis was recently supported by the detection of large quantities of cyanonaphthalene, cyanopyrene and cyanocoronene compounds in the Taurus Molecular Cloud 1 region.\cite{mcguire2021detection,Wenzel:2025ab,Wenzel:2025aa}\par

Such a detection raises questions about the stability of small PAHs in the interstellar medium, due to their likely vulnerability to VUV photons of energies up to 13.6 eV.\cite{Jochims:1999fk} Shortly after their existence was theoretically postulated, it was suggested that small PAHs might undergo a specific radiative relaxation process called recurrent fluorescence (RF), also known as Poincaré fluorescence,\cite{leger1988predicted} which would enhance their stability. According to the transient heating mechanism,\cite{Sellgren:1983aa} PAH molecules can be excited to high electronic states upon absorption of visible or UV radiation. Individual molecules can then rapidly undergo internal conversion (IC), transferring electronic excitation into vibrational energy in the electronic ground state. Subsequently, relaxation can proceed further through the emission of IR photons, thereby causing the AIBs.\par
\begin{figure}
\centering
\includegraphics{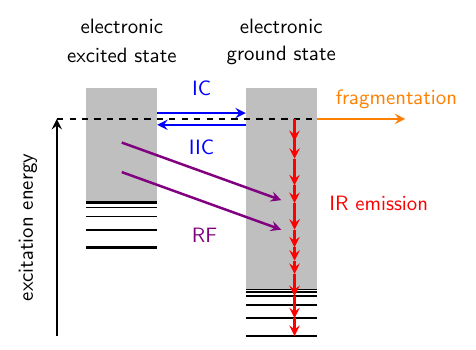}
\caption{Jablonski diagram showcasing  some of the most important processes involved in the stochastic heating mechanism: initial excitation, IR photon emission, fragmentation, internal conversion (IC), inverse internal conversion (IIC) and recurrent fluorescence (RF).}
\label{fig:jabdiag}
\end{figure}
Among the various possible relaxation mechanisms taking place upon transient heating, which notably include fragmentation, RF can occur when sufficient energy remains in the vibrational modes of the molecule to excite a low-lying electronic excited state through inverse internal conversion (IIC).\cite{Karny:1979aa} Subsequent relaxation to the electronic ground state can occur via electronic fluorescence in the near-IR or visible regions (see Fig.~\ref{fig:jabdiag}). This alternative mechanism is not observed in dense media, in which collisions with the surrounding environment dissipate energy faster than the characteristic IIC timescale, but it becomes relevant under low densities. For example, RF was directly observed  over time scales of a few $\mu$s in the Zare group\cite{Karny:1979aa} as delayed electronic luminescence following multiphoton IR photon absorption of small molecules in collision-free conditions. In the astrochemical context, RF was proposed by Léger and coworkers\cite{leger1988predicted} as a possibly important mechanism occurring in PAH molecules in the ISM over timescales that can be significantly extended up to a few ms. More recently, RF was suggested to occur in neutral carbon clusters, a potentially important compound in the ISM, possibly contributing to the near-IR emission continuum.\cite{Lacinbala:2023aa}\par

A large number of dedicated experiments on the relaxation kinetics of PAHs have relied on ion traps\cite{boissel1997fragmentation,saito2020direct,kusuda2024detection} and storage rings.\cite{Martin:2013wt,Martin:2015aa,Martin:2017uu,Bernard:2017vf,Bernard:2023aa,Lee:2023aa,stockett2023efficient,bull2025radiative} These setups replicate ISM-like conditions by providing a low-density environment, enabling the study of collisional-free relaxation of hot PAH cations over long timescales reaching tens of milliseconds and more. The internal energy remaining in the molecular ions is typically monitored by measuring the occurrence of dissociation byproducts during the relaxation. In particular for PAHs, in the excitation energy range relevant for astrochemistry ($<13.6$~eV), such byproducts usually consist of atomic hydrogen and acetylene.\cite{West:2012aa,West:2014ug,West:2018aa} These experiments have revealed that multiple relaxation mechanisms, including RF, compete with each other to dissipate energy, thereby providing an indirect access to the RF process.\par

The RF-delayed emission mechanism was also directly observed by Saito and coworkers for naphthalene and antracene~\cite{saito2020direct,kusuda2024detection} in the wavelength region between 500 and 850 nm, using various bandpass filters. These authors observed an emission profile comparable to the absorption spectrum in the short-wavelength side but with a stronger component in the long-wavelength side, whose origin remains unclear.\par

Already from its original conceptualization, RF has been identified as a statistical process and its theoretical framework has seen little advancement since the pioneering work of Nitzan and Jortner.\cite{Nitzan:1978aa,Nitzan:1979aa} These authors described RF as a vertical transition, using the Condon approximation applied to a microcanonical statistical population in the electronic excited state. This model has been applied to individual molecules\cite{Chandrasekaran:2014aa,Lacinbala:2022uj,Pedersen:2025aa} and was shown to lead to blackbody-like emission spectra when applied to a broad molecular sample.\cite{Lacinbala:2023ab} However, in many applications, this simple model has been recognized to significantly underestimate the overall RF cooling rates. \cite{Martin:2013wt,Stockett:2020uk,Stockett:2025ab} The model was recently improved to include the Herzberg-Teller (HT) contribution\cite{Herzberg:1966aa} in the transition dipole moment between the electronic ground and excited states.\cite{stockett2023efficient} In some particular cases, the HT coupling is known for significantly modulating the spectroscopic response,\cite{Kundu:2022aa} and in the context of RF, it was notably shown that the HT coupling can increase the stability of cyanonaphthalene by strongly increasing the oscillator strength of the lowest electronic transition. However, in the approach led by Stockett and coworkers,\cite{stockett2023efficient} the remaining vibrational energy in the electronic excited state was not considered and the effects of the vibrational structure of both electronic states were still ignored, hindering a more complete interpretation of the spectra measured by Saito {\em et al.}\cite{saito2020direct,kusuda2024detection}

The inclusion of the molecular vibrational structure in the modeling of absorption and emission spectroscopy experiments can be achieved by  explicitly computing the Franck-Condon (FC) factors between the various vibronic states.\cite{Weber:2003aa,Huh:2012kx} This method is applicable for harmonic potential energy surfaces, with or without the Duschinsky rotation effects, as well as for anharmonic potential energy surfaces. Unfortunately, it is computationally tractable only for moderately large systems and for limited numbers of initial states. Alternatively, a more global method for determining the spectroscopic signature, based on the transition dipole moment time correlation function, is particularly efficient for harmonic systems at thermal equilibrium and can readily account for the Duschinsky rotation.\cite{Yan:1986aa,Ianconescu:2004aa,Baiardi:2013aa} To date, neither approach has been applied to the study of RF processes.\par

\begin{figure}
    \centering
\includegraphics{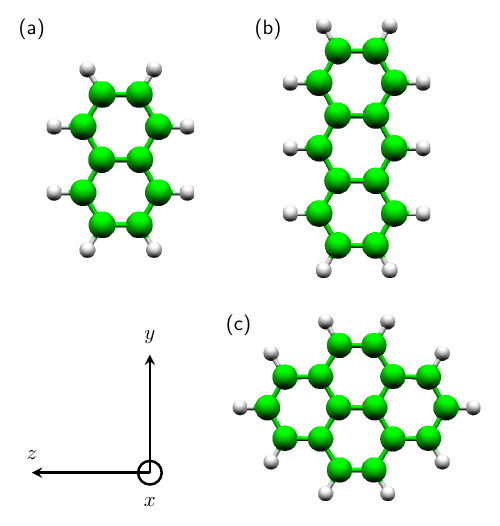}
     \caption{PAH cations studied in this work. (a) Naphthalene; (b) anthracene; (c) pyrene, together with the reference frame used to identify appropriate symmetries.}
    \label{fig:molecules}
\end{figure}

Many archetypal PAHs used to model the AIBs exhibit a high symmetry, and such is the case for the naphthalene (Np), anthracene (An), and pyrene (Py) cations depicted in Fig.~\ref{fig:molecules}, which all fall into the $D_{2h}$ point group. The electronic states of these three cations have been extensively studied experimentally using photoelectron (PE)\cite{,boschi1972photoelectron,boschi1972photoelectron-bis,bally1998electronic,Coropceanu:2002aa,Sanchez-Carrera:2006aa,mayer2011threshold,West:2012aa,West:2014aa,West:2014ug} and absorption\cite{Salama:1991aa,szczepanski1993electronic,vala1994electronic,Pino:1999bh,Romanini:1999aa,Sukhorukov:2004aa,Biennier:2004aa,Meyer:2021aa} spectroscopies. Symmetry plays a crucial role in the optical response of molecular systems, and can alone forbid excitations to specific electronic states.\cite{hirata2003time}  In the (fairly large) Np$^+$, An$^+$, and Py$^+$ cations, the lowest-lying electronic excited state, commonly noted as D$_1$, is most likely to be populated through the IIC mechanism in the statistical limit. However, D$_1$ is also a dark state and thus does not contribute to the emission spectrum, which explains why the current modeling of RF emission and its associated rate constants in such molecules have usually ignored this state, focusing exclusively on the second (bright) electronic excited state, D$_2$.\cite{boissel1997fragmentation,Martin:2015aa,saito2020direct,Lee:2023aa} However,  at finite internal energy, the excitation of vibrational modes with ungerade symmetry breaks the inversion symmetry of the molecules, possibly leading to a finite transition dipole moment, which could in turn enable the HT effect, confirming the importance of such processes to explain the stability of non inversion-symmetric molecules such as cyanonaphthalene.\cite{stockett2023efficient} To the best of our knowledge, in the case of inversion-symmetric molecules like those depicted in Fig.~\ref{fig:molecules}, these effects have only been considered in a very recent study,\cite{Subramani:2025aa} where RF was modeled as an average of vertical transitions from individual geometries extracted from \emph{ab initio} molecular dynamics simulations. This approach naturally includes the symmetry breaking due to presence of internal energy but does not include all the HT effects, which manifest even in the absence of internal energy.

The present work pursues two main objectives. First, a new theoretical framework for relaxation by recurrent fluorescence emission in high-dimensional molecular systems is introduced, based on the harmonic approximation and explicitly accounting for both HT couplings and vibrational progressions. The approach relies on the calculation of the transition dipole moment time autocorrelation function, and a practical but accurate solution involving a canonical representation of the true microcanonical energy distribution. Second, the model is applied to quantify the contribution of symmetry-forbidden transitions in the inversion-symmetric Np$^+$, An$^+$ and Py$^+$ cationic PAHs to their RF emission spectrum and the associated rate constants. 

The article is organized as follows. In Sec.~\ref{sec:theory}, we present the theoretical model, from a general perspective first, before proceeding to a practical solution based on harmonic models. The results are introduced in Sec.~\ref{sec:results}, discussed in the light of the importance of the D$_1$ states and their dependence on the PAH itself, and compared with existing experiments. Finally, conclusions are drawn in Sec.~\ref{sec:conclusions}, together with some possible future extensions. More technical appendices are also included to relieve the main text.

\section{Theoretical model for RF relaxation}
\label{sec:theory}
\subsection{Recurrent fluorescence differential rate constant}
A nonrotating molecule is described by its electronic ground state $\ket{g}$ and one electronic excited state denoted by $\ket{e}$.  From this two-state model, the generalization to an arbitrary number of electronic states is straightforward.
For both states $\ket{g}$ and $\ket{e}$, the vibrational dynamics is described by appropriate Hamiltonians $H_g$ and $H_e$, respectively. Within the Born-Oppenheimer approximation, the total Hamiltonian $H$ of the system is expressed as
\begin{equation}
    H = H_g\ket{g}\bra{g} + H_e\ket{e}\bra{e}.
\end{equation}
For a molecule with internal energy $E$ relative to the fundamental vibronic state, the dimensionless differential RF rate constant is given by
\begin{equation}
    k^{\text{RF}}(\omega;E) = \sum_{i,f}P_i(E)A_{if}\delta(\omega-\omega_{if}),
    \label{eq:rfrate1}
\end{equation}
where $i$ and $f$ are the initial and final vibronic states, respectively, $P_i(E)$ the occupation probability of state $i$ and $\omega_{if} = \omega_i - \omega_f$ is the transition frequency.  In Eq.~(\ref{eq:rfrate1}), the Einstein coefficient $A_{if}$ for spontaneous emission is expressed as
\begin{equation}
    A_{if} = \frac{\omega_{if}^3}{3\pi\epsilon_0\hbar c^3}\left|\vect{d}_{if}\right|^2,
\end{equation}
where $\vect{d}_{if}= \bra{i} \vect{d} \ket{f}$ is the transition dipole moment between the initial and final states. Using the expression of the Fourier transform of the delta function, the RF differential rate constant can be rewritten as
\begin{equation}
    k^{\text{RF}}(\omega;E) = \frac{\omega^3}{3\pi\epsilon_0\hbar c^3}\int^{+\infty}_{-\infty}\frac{\ud t}{2\pi}\ \tr\left[\rho(E)\vect{d}(0)\cdot \vect{d}(t)\right]e^{\ii\omega t},
     \label{eq:rfrate2}
\end{equation}
where the trace acts over all vibronic states. In Eq.~(\ref{eq:rfrate2}), the transition dipole moment is given by 
\begin{equation}
\vect{d} = \vect{d}_{eg} \ket{e} \bra{g} + \vect{d}_{ge} \ket{g} \bra{e},
\end{equation}
where $\vect{d}_{eg}$ and $\vect{d}_{ge}$ are operators acting in the vibrational coordinates space. Within a microcanonical statistical model, the density matrix $\rho(E)$ is given by
\begin{equation}
\rho(E) = \frac{ \delta(H-E)}{\Omega(E)},
\end{equation}
with $\Omega(E) = \tr \left[ \delta(H -E) \right]$ the density of vibronic states. The density matrix can be expanded over the two electronic states as
\begin{equation}
\rho(E) = \rho_g(E) \ket{g}\bra{g} + \rho_e(E) \ket{e}\bra{e},
\end{equation}
where the vibrational density matrices $\rho_g(E)$ and $\rho_e(E)$ are given by
\begin{equation}
\rho_i(E) = \frac{\delta(H_i -E )}{\Omega_g(E) + \Omega_e(E)}, \quad i=g,e.
\end{equation}
The densities of vibrational states $\Omega_g(E)$ and $\Omega_e(E)$ in each electronic state are formally expressed using the trace over vibrational states $\tr_v$,
\begin{equation}
\Omega_i(E) = \tr_v\left[ \delta(H_i - E)\right] \quad i=g,e.
\end{equation}
Therefore the total probability to be in the electronic excited state is given by
\begin{equation}
P_e(E)= \tr_v\left[\rho_e(E)\right]= \frac{\Omega_e(E)}{\Omega_g(E) + \Omega_e(E)}.
\end{equation}
We next assume that fluorescence can only occur from the excited electronic state to the electronic ground state. The RF differential rate constant is thus written as
\begin{equation}
    k^{\text{RF}}(\omega;E) = \frac{\omega^3}{3\pi\epsilon_0\hbar c^3}\int^{+\infty}_{-\infty}\frac{\ud t}{2\pi}\ \tr_v[\rho_e(E) e^{-\ii H_e t/\hbar} \vect{d}_{eg}e^{\ii H_gt/\hbar}\vect{d}_{ge}] e^{\ii\omega t}.
     \label{eq:rfrate3}
\end{equation}
Solving the correlation function inside the integral of Eq.~(\ref{eq:rfrate3}) directly in the microcanonical ensemble requires to compute explicitly the trace over vibrational states. This operation is computationally expensive, even under the harmonic approximation, and only feasible for systems with a very small number of degrees of freedom, especially when accounting for the Duschinsky effect that couples the various normal modes together. However, and following earlier suggestions,\cite{Ianconescu:2004aa,Baiardi:2013aa} the calculation becomes significantly more tractable for harmonic systems governed by the {\em canonical}\/ ensemble. For very large systems, the equivalence between the canonical and microcanonical ensembles justifies this approximation. However, this is less straightforward for systems with a limited number of degrees of freedom such as Np$^+$.

To proceed, the energy within each state is then assumed to follow a Boltzmann distribution at an effective canonical temperature $T_{\rm eff}(E)$, rather than a delta-peaked distribution that is characteristic of the true microcanonical distribution, and while keeping this microcanonical distribution for determining the occupancy probability of the electronic ground and excited states. At any arbitrary energy $E_0$, and for any electronic state $i$, the microcanonical temperature $T_\mu$ at thermal equilibrium is defined from the density of vibrational states as
\begin{equation}
\beta_{\mu}(E_0) = \frac{1}{\boltz T_\mu} = \frac{\partial\ln \Omega_i}{\partial E} \Bigg|_{E=E_0}.
\label{eq:microtemp}
\end{equation}

The probability distribution $p_i(E)$ of energies in the microcanonical ensemble is a delta peak at energy $E_0$, but becomes broad once in the presence of a thermostat. In the canonical ensemble at inverse temperature $\beta=1/k_{\rm B}T$, and for the very same electronic state, the corresponding probability distribution reads
\begin{equation}
p_i^{(T)}(E) \propto \Omega_i(E) e^{-\beta E},
\label{eq:approx}
\end{equation}
where the normalizing prefactor is the inverse of the canonical partition function.

For any given microcanonical energy $E_0$, several choices are available for the effective canonical temperature $T_{\rm eff}(E_0)$
based on the above probability distributions. One possibility consists in adjusting the maximum of the canonical distribution, $\max_E \{ p_i^{(T)}(E) \}$, to the microcanonical energy value, thereby providing a first value for $T_{\rm eff}$ denoted as $T_{\rm max}$. By differentiation of Eq.~(\ref{eq:approx}) it can be straightforwardly shown that the maximum of the distribution occurs when $T_{\rm max}=T_\mu$. 

Alternatively, instead of the maximum, the average $\langle E\rangle$ of the distribution can be adjusted to match the microcanonical energy, the corresponding equation $\langle E\rangle(T)=E_0$ being always well defined owing to the monotonic character of $\langle E\rangle(T)$. Unlike $T_{\rm max}$, which is rooted in microcanonical statistics, the effective temperature $T_{\rm av}$ resulting from matching $\langle E\rangle$ to the required $E_0$ has a canonical perspective.

Except for classical harmonic systems in the large size limit, $T_{\rm max}$ and $T_{\rm av}$ generally differ from one another. However, and as will be shown below, this difference is rather small already for the present systems, and the choice of $T_{\rm av}$ as the effective canonical temperature turns out to be slightly more accurate than $T_{\rm max}$.

Denoting simply $\beta_{\rm eff}=1/k_{\rm B}T_{\rm eff}$ the inverse effective canonical temperature, the differential RF rate constant is next rewritten as
\begin{equation}
  k^{\text{RF}}(\omega;E) = \frac{\omega^3}{3\pi\epsilon_0\hbar c^3}P_e(E) \int^{+\infty}_{-\infty}\frac{\ud t}{2\pi}\   \frac{1}{Z_e(T_{\text{eff}})} C_{ge}(-t,t-\ii\beta_{\text{eff}}\hbar ) e^{\ii\omega t},
  \label{eq:rfratecstfinal}
\end{equation}
where $Z_e(T)=\tr[e^{-\beta H_e} ] $ is the equilibrium partition function in the electronic ground state and where the correlation function $C_{ge}(\tau_g,\tau_e)$ reads
\begin{equation}
C_{ge}(\tau_g,\tau_e) = \tr_v \left[  e^{-\ii H_g \tau_g/\hbar } \vect{d}_{ge} e^{-\ii H_e \tau_e/\hbar } \vect{d}_{eg} \right].
\label{eq:correl1}
\end{equation}
From this expression, the total RF rate constant is derived as
\begin{equation}
k^{\text{RF}}_{\text{tot}}(E) = \int_{0}^\infty \!  \ud \omega  \ k^{\text{RF}}(\omega;E).
\label{eq:totalRF}
\end{equation}

At any physical temperature $T=1/k_{\rm B}\beta$, the equilibrium absorption cross-section can be obtained using a similar correlation function
\begin{equation}
\sigma^{\text{abs}}(\omega;T) = \frac{\pi \omega}{3\hbar \epsilon_0 c} \left(1 - e^{-\beta\hbar\omega} \right) \int^{+\infty}_{-\infty}\frac{\ud t}{2\pi}\  \frac{1}{Z_g(T)} C_{ge}\left(-t - \ii \beta \hbar , t\right) e^{\ii \omega t},
\label{eq:absspectrum}
\end{equation}
where $Z_g(T)=\tr[e^{-\beta H_g} ] $ is the equilibrium partition function in the electronic ground state. The same formalism provides the PE spectrum between the electronic ground state $g$ of the neutral molecule and any arbitrary electronic state $e$ (which can also be the electronic ground state) of the corresponding cation. For this type of spectroscopy, since the transition probability is generally not accessible, it is common practice to assume a vertical transition with an arbitrary transition dipole moment.\cite{Kouppel:1984aa} The spectral intensity is then proportional to
\begin{equation}
{\cal I}^{\text{PE}}(\omega;T) = \int^{+\infty}_{-\infty}\frac{\ud t}{2\pi}\  \frac{1}{Z_g(T)} C^{\text{PE}}_{ge}\left(-t - \ii \beta \hbar , t\right) e^{\ii \omega t},
\label{eq:PE}
\end{equation}
where $C^{\text{PE}}_{ge}(\tau_g,\tau_{e})$ is given by
\begin{equation}
C^{\text{PE}}_{ge}(\tau_g,\tau_{e}) = \tr_v \left[  e^{-\ii H_g \tau_g/\hbar } e^{-\ii H_{e} \tau_e/\hbar } \right].
\label{eq:correlPE}
\end{equation}
The calculation of the correlation function from Eqs.~(\ref{eq:correl1}) and (\ref{eq:correlPE}) requires the knowledge of the eigenvalues and eigenstates of the vibrational Hamiltonians $H_g$  and $H_e$.  In the general anharmonic case, the diagonalization step and the sum-over-states numerical evaluation of the correlation function can only be achieved for small systems. However, the problem can be largely simplified under the harmonic approximation.

\subsection{Harmonic model}

Following Refs.~\citenum{Ianconescu:2004aa} and \citenum{Baiardi:2013aa}, the harmonic approximation is used to obtain the differential RF rate constants. Within this approximation, the potential energy surface is expanded up to second order around the electronic ground state equilibrium position and $H_g$ is expressed as a function of the $n=3N-6$ normal mode coordinates $q_{g,i}$ and the conjugated momenta $p_{g,i}$, the total number of atoms being denoted as $N$. The Hamiltonian is then written as
\begin{equation}
H_g = \sum_i^n \frac{1}{2} \left( p_{g,i}^2+\omega_{g,i}^2 q_{g,i}^2\right),
\label{eq:harmonicg}
\end{equation}
where $\omega_{g,i}$ denotes the set of ground state vibrational frequencies. In the electronic excited state, a similar expansion can be performed around the corresponding equilibrium position and the Hamiltonian is expressed as a function of the  normal mode coordinates $q_{e,i}$ and the conjugated momenta $p_{e,i}$ as
\begin{equation}
H_e = \sum_i^n \frac{1}{2} \left( p_{e,i}^2+\omega_{e,i}^2 q_{e,i}^2\right) + \hbar \omega_{eg},
\label{eq:harmonice}
\end{equation}
where $\omega_{e,i}$ now denotes the set of electronic excited state vibrational frequencies, with $\omega_{eg}$ being the adiabatic transition frequency connecting the two electronic states. Both sets of normal modes coordinates are linearly connected to each other,
\begin{equation}
    \vect{q}_e = \vect{J}\vect{q}_g+\vect{K},
    \label{eq:duschinsky}
\end{equation}
where $\vect{q}_g = \left(q_{g,1},q_{g,2},\dots,q_{g,n}\right)$ and $\vect{q}_e = \left(q_{e,1},q_{e,2},\dots,q_{e,n}\right)$ are two $n$-tuples containing the electronic ground and excited states normal mode coordinates. In Eq.~(\ref{eq:duschinsky}), $\vect{J}$ is the Duschinsky rotation matrix defined as the projection of the electronic ground state normal mode coordinates on the electronic excited state normal mode coordinates and $\vect{K}$ is the displacement vector corresponding to the difference in equilibrium positions in the electronic ground and excited states, expressed in the basis set of the electronic excited state normal mode coordinates. The detailed expressions of the matrix $\vect{J}$ and the vector $\vect{K}$ as a function of the electronic states potential energy surface parameters are given in Appendix~\ref{sec:app_duschinsky}.

Within the harmonic approximation, the canonical partition function in the electronic excited state is written
\begin{equation}
Z_e(T_{\text{eff}}) = \prod_{i=1}^n \frac{1}{2\sinh\beta_{\text{eff}}\hbar\omega_{e,i}/2},
\end{equation}
and a similar expression for the electronic ground state. The transition dipole moment between the electronic ground and excited states can also be expanded linearly around the electronic ground state geometry as 
\begin{equation}
    \vect{d}_{eg}=\vect{d}_{ge} \approx \vect{d}^{0}_{g}+\sum_i\vect{d}^{1}_{g,i}q_{g,i} = \vect{d}^{0}_{g} + \vect{d}^1_g \vect{q}_g, \label{eq:transdipoleground}
\end{equation}
where the three-dimensional vector $\vect{d}^{0}_{g}$ and the $3\times n$  matrix $\vect{d}^1_g$ give the Franck-Condon and Herzberg-Teller contributions to the vibronic spectra, respectively. Alternatively, the transition dipole moment can be expanded linearly around the electronic excited state equilibrium geometry as
\begin{equation}
    \vect{d}_{eg} \approx \vect{d}^{0}_{e} + \vect{d}^1_e \vect{q}_e.
    \label{eq:transdipoleexcited}
\end{equation}
Under this linear approximation, Eqs.~(\ref{eq:transdipoleground}) and (\ref{eq:transdipoleexcited}) are thus connected to one another through
\begin{align}
    &\vect{d}^0_{g} = \vect{d}^0_{e}+\vect{d}^1_{e}\vect{K},\\
    &\vect{d}^1_{g} = \vect{d}^1_{e} \vect{J}.\label{eq:compdipole}
\end{align}
In the expression of the differential rate constant [Eq.~(\ref{eq:rfratecstfinal})], the correlation function $C_{ge}(-t,t-\ii\beta_{\text{eff}} \hbar)$  can be easily computed under these approximation and in the $t\rightarrow 0$ limit. From the definition of the time correlation function, it reduces to the average squared transition dipole moment at temperature  $T_{\text{eff}}$ in the electronic excited state 
\begin{equation}
 \frac{1}{Z_e(T_{\text{eff}})} C_{ge}(0,-\ii\beta_{\text{eff}}\hbar) =  \left|\vect{d}_e^0\right|^2
         +\sum_{k} \frac{\hbar}{2\omega_{e,k}} \left|\vect{d}^1_{e,k}\right|^2 \coth\left(\beta_{\text{eff}} \hbar \omega_{e,k}\right).
\label{eq:tzl}
\end{equation}
From Eq.~(\ref{eq:tzl}), the correlation function controlling the emission spectrum contains two sets of contributions: the Franck-Condon contribution given by the transition dipole moment at the electronic excited state equilibrium geometry and the HT contributions arising from each normal mode. The FC contribution can be determined from the oscillator strength $f^{\text{FC}}_{e}$, defined here as
\begin{equation}
f^{\text{FC}}_{e} = \frac{2m_e}{3\hbar e^2} \omega_{ge} \left|\vect{d}_e^0\right|^2.
\label{eq:fFC}
\end{equation}
The temperature-dependent HT contribution $f^{\text{HT}}_{e}$ is obtained from the corresponding oscillator strength 
\begin{equation}
 f^{\text{HT}}_{e}(T_{\text{eff}}) = \sum_k f^{\text{HT}}_{e,k}(T_{\text{eff}}) = \sum_k \frac{m_e \omega_{ge}}{3 e^2\omega_{e,k}} \left|\vect{d}^1_{e,k}\right|^2 \coth\left(\beta_{\text{eff}} \hbar \omega_{e,k}\right).
 \label{eq:fHT}
\end{equation}
Note that when the electronic transition is forbidden by symmetry, the transition dipole moment at the equilibrium geometry vanishes ($\vect{d}_e^0=\vect{0}$). However, in general the normal mode contribution will not vanish for any ungerade mode that breaks the inversion symmetry of the molecule, even at vanishing temperature, due to the delocalized nature of the wavefunction around the equilibrium position.\par
To compute the RF rate constants [Eq.~(\ref{eq:rfratecstfinal})] and the absorption [Eq.~(\ref{eq:absspectrum})] and PE [Eq.~(\ref{eq:PE})] spectra, the correlation function $C_{ge}(\tau_g,\tau_e)$ needs to be evaluated for all possible values of the complex variables $\tau_g$ and $\tau_e$. A closed-form expression for this correlation function, excluding the HT contribution, was derived by Ianconescu and Pollak.\cite{Ianconescu:2004aa} Baiardi {\em et al.} later extended this expression to include the HT contribution.\cite{Baiardi:2013aa} Both studies arrive at essentially the same result, albeit using different notations. In Appendix~\ref{sec:app_correl}, we present the final expression for the correlation function $C_{ge}(\tau_g,\tau_e)$, including the HT contribution, as used in our calculations. The corresponding derivation, following the notations of Ianconescu and Pollak,\cite{Ianconescu:2004aa} is also provided. The final expression of Eq.~(\ref{eq:correlfinal}) was employed to compute the various spectra through Eqs.~(\ref{eq:rfratecstfinal}), (\ref{eq:absspectrum}), and (\ref{eq:PE}).
\subsection{Computational details}

The correlation function [Eq.~(\ref{eq:correlfinal})] used in the various spectral calculations [Eqs.~(\ref{eq:rfratecstfinal}), (\ref{eq:absspectrum}), and (\ref{eq:PE})] requires  the knowledge of the harmonic expansion of the potential energy surfaces  for the electronic ground and excited states around their respective equilibrium positions [Eqs.~(\ref{eq:harmonicg}) and~(\ref{eq:harmonice})] as well as the linear expansion of the transition dipole moment between the two states around the electronic ground or excited states equilibrium geometries [Eqs.~(\ref{eq:transdipoleground}) or~(\ref{eq:transdipoleexcited})].
All these parameters were obtained using quantum chemistry calculations. In the past, a large number of studies conducted at various theoretical levels have been devoted to the description of PAHs electronic excited states.\cite{szczepanski1993electronic,vala1994electronic,bally1998electronic,Negri:1994aa,Negri:1997zr,Hirata:1999aa,hirata2003time,Dierksen:2004ab,Dierksen:2004aa,Malloci:2007xq,Malloci:2011aa,Ghanta:2011aa,Boggio-Pasqua:2019aa,Knysh:2024aa} For low-lying electronic excited states, time-dependent density-functional theory (TD-DFT)\cite{Casida:1998aa,Casida:2012aa} has been shown to give reliable results for PAHs with various charge states.\cite{hirata2003time,Dierksen:2004ab,Dierksen:2004aa,Friha:2013uq,Boggio-Pasqua:2019aa,Knysh:2024aa}
In particular, for neutral PAHs, TD-DFT was shown to correctly describe weakly dipole-allowed and dipole-forbidden transitions in the calculation of vibronic absorption spectra simulated within the Herzberg-Teller approximation.\cite{Dierksen:2004ab,Dierksen:2004aa}

The structure and properties of the electronic ground state denoted as $S_0$ and $D_0$ for the neutral and cationic molecules, respectively, were obtained using density-functional theory (DFT) calculations with the $\omega$B97xD functional~\cite{Chai:2008nr} combined with the cc-pVTZ basis set.\cite{Dunning:1989aa,Kendall:1992xy} 
Electronic excited states properties of cations were determined using TD-DFT at the same level of theory as for the electronic ground states. The same functional and basis set were used recently to describe RF fluorescence in naphthalene and azulene.\cite{Lee:2023aa} All electronic structure calculations were performed using the Gaussian16 quantum chemistry software package.\cite{g16} 

After locating the minimum geometries in the electronic ground and excited states using DFT and TD-DFT, respectively, the Hessian matrices in mass-scaled Cartesian coordinates were extracted and diagonalized to determine the harmonic frequencies and associated normal mode coordinates. The minimum geometries and normal mode coordinates were subsequently used to compute the displacement vector $\vect{K}$ and the Duschinsky rotation matrix $\vect{J}$ (see Appendix~\ref{sec:app_duschinsky}). From the TD-DFT calculation, we also extracted the transition dipole moment at the minimum geometry of the excited electronic state, $\vect{d}^0_e$, as well as its first derivative in Cartesian coordinates. These derivatives were then converted to normal mode derivatives to obtain the parameters $\vect{d}^1_{e,k}$ [Eq.~(\ref{eq:transdipoleexcited})].

Alternatively, and as argued by Baiardi and coworkers,\cite{Baiardi:2013aa} the electronic ground state minimum geometry can also be used to extract directly the transition dipole moment parameters $\vect{d}^1_g$. In the linear approximation, $\vect{d}^1_e$ and $\vect{d}^1_g$ are related to one another through Eq.~(\ref{eq:compdipole}). To assess the validity of the linear approximation, we introduce the following dimensionless quantity 
\begin{equation}
\varepsilon = \frac{\sum_{k=1}^n \left| \vect{d}^1_{e,k} -  \sum_i \vect{d}^1_{g,i} S_{k,i} \right|^2 } { \sum_{k=1}^n \left| \vect{d}^1_{e,k}\right|^2},
\end{equation}
which vanishes only if the linear expansion of the dipole moment is exact. We computed $\varepsilon$ for Np$^+$ and for both electronic transitions. In the case of the D$_0$ $\rightarrow$ D$_{2}$ transition, very small variations in the transition dipole moment between the electronic ground state geometry and the electronic excited state geometry are obtained, with $\varepsilon =4$\%. In contrast, for the D$_0$ $\rightarrow$ D$_{1}$ transition, the variations are more pronounced with $\varepsilon =23$\%. Nevertheless, and while the results presented below may quantitatively depend on the choice of reference, the qualitative conclusions should generally hold.

Microcanonical properties depend on the density of vibrational modes in the various electronic states. Within the harmonic approximation, the densities of vibrational states were computed using the Stein-Rabinovitch counting algorithm,\cite{stein1973accurate} which is more accurate than the Beyer-Swinehart method\cite{Beyer:1973aa} and does not introduce rounding errors to the mode energies. The densities of states were computed on a regular grid with an energy bin of $\Delta E = 100~\cm$ and up to a value of $E_{\text{max}}=110000~\cm$. Finally, the time correlation functions of Eq.~(\ref{eq:correlfinal}) were computed over 16.7 ps on a regular time grid of 5,001 points with a timestep of 0.167 fs using a in-house code. 
Values of the correlation function at negative times were obtained by complex conjugation, exploiting the property $C_{ge}(-\tau_g^*,-\tau_e^*) = C_{ge}^*(\tau_g,\tau_e)$, thereby increasing the total number of points to 10001. The various spectra defined by Eqs.~(\ref{eq:rfratecstfinal}), (\ref{eq:absspectrum}) and (\ref{eq:PE}) were then determined using fast Fourier transform, yielding a resolution of $20~\cm$. In practice, to impose a vanishing time-correlation function at long time scales, the time correlation functions were convoluted with a Gaussian function of fixed width. Gaussian functions with a full width at half maximum (FWHM) of $400~\cm$ were used for electronic absorption and PE spectra, a width of $50$~\cm\ being used for RF spectra. 
\section{Results}
\label{sec:results}
\subsection{Electronic structure of the cations}
For all Np$^+$, An$^+$, and Py$^+$ cations, the two lowest electronic excited states D$_1$ and D$_2$ were considered in the analysis. For the pyrene cation, D$_3$ was also included due to its energetic proximity with D$_2$. For each electronic excited state, the geometry was locally optimized, starting from the electronic ground state equilibrium configuration. After optimization, the $D_{2h}$ point group is preserved in all electronic excited states. Table~\ref{tab:electronicstates} lists the main energetic features of the various electronic excited states, namely their adiabatic relative energy with respect to the neutral molecule in the electronic ground state, and the transition energy between the vibrational ground state of the electronic states of the cation relative to the ground vibrational and electronic ground state of the neutral molecule. This energy includes the zero-point contribution in the harmonic approximation and can be directly compared to photoelectron spectroscopy measurements when the origin band is well isolated from the other vibronic bands.\cite{bally1998electronic,szczepanski1993electronic,boschi1972photoelectron,boschi1972photoelectron-bis,mayer2011threshold} 
Our quantum chemical results are very similar to earlier TD-DFT calculations.\cite{Hirata:1999aa,Malloci:2004aa,Malloci:2007xq} Compared to experimental values, the error on the position of the cationic electronic states with respect to the neutral electronic ground state ranges from 0.01~eV (D$_1$ state of Py$^+$) to 0.31~eV (D$_0$ state of An$^+$), and it is noteworthy that the largest errors occur for the electronic ground states. Overall, our calculations tend to underestimate the transition energies with respect to experimental data, except for the D$_3$ state of Py$^+$ which is slightly overestimated by 0.06 eV.
\par
\begin{table*}
\begin{tabular*}{\linewidth}{@{\extracolsep{\fill}}llcccccc}
\hline\hline
Molecule & State  & relative energy (eV) & \multicolumn{2}{c}{transition  energy (eV) } & $f^{\text{FC}}_{e} (\times 10^3)$ & \multicolumn{2}{c}{$f^{\text{HT}}_{e} (\times 10^3)$ }  \\
\cline{4-5}\cline{7-8}
& &   & calc. & exp. &  & 0 K& 2000 K \\
\hline
 Np & S$_0$ (X$^1$A$_g$) & ~~0 & ~~0 & ~~0 & -- & -- & -- \\
 \multirow{3}{*}{Np$^+$} & D$_0$ (X$^2$A$_u$) & 7.92 & 7.91 & 8.15$^a$ & -- & -- & -- \\
& D$_1$ (A$^2$B$_{3u}$) & 8.74 & 8.76 & 8.88$^a$ & $\times$ & 0.6596 & 0.9503 \\
& D$_2$ (B$^2$B$_{2g}$) & 9.81 & 9.79 & 10.08$^a$ & 82.12 & 0.7252 & 1.926 \\
\hline
An & S$_0$ (X$^1$A$_g$) & ~~0 & ~~0 & ~~0 & -- & -- & --  \\
\multirow{3}{*}{An$^+$}  & D$_0$ (X$^2$B$_{2g}$) & 7.15 & 7.16 & 7.47$^b$ & -- & -- & -- \\
& D$_1$ (A$^2$B$_{1g}$) & 8.53 & 8.52 & 8.57$^b$ & $\times$ &  4.745 & 7.641 \\
& D$_2$ (B$^2$A$_{u}$) & 9.03 & 9.05 & 9.23$^b$ & 164.0 & 1.113 & 1.571 \\
\hline
Py & S$_0$ (X$^1$A$_g$) & ~~0 & ~~0 & ~~0 & -- & -- & -- \\
\multirow{4}{*}{Py$^+$}   & D$_0$ (X$^2$B$_{1g}$) & 7.19 & 7.19 & 7.41$^c$ & -- & -- & -- \\
& D$_1$ (A$^2$B$_{2g}$) & 8.25 & 8.26 & 8.26$^c$ & $\times$ & 1.243 & 1.817 \\
& D$_2$ (B$^2$B$_{3u}$) & 8.94 & 8.93 & 9.00$^c$ & 34.34 & 4.565 & 10.07 \\
& D$_3$ (C$^2$A$_{u}$) & 9.32 & 9.35 & 9.29$^c$ & 58.67 & 15.43 & 18.44 \\
\hline\hline
\multicolumn{6}{l}{$^a$ Ref.\citenum{bally1998electronic}, $^b$ Refs.~\citenum{boschi1972photoelectron} and~\citenum{szczepanski1993electronic}, $^c$ Ref. \citenum{boschi1972photoelectron-bis}} \\
\end{tabular*}
\caption{Spectroscopic notation, relative energy, transition  energy and transition dipole moment contributions of the various electronic states considered in this study. The symbol $\times$ is used to indicate a symmetry-forbidden transition.}
\label{tab:electronicstates}
\end{table*}
In order to evaluate the quality of the potential energy surfaces in the electronic ground and excited states with the present (TD-)DFT methods, the theoretical PE spectra were determined for the three cations using Eq.~(\ref{eq:PE}). To refine the model further, the positions of the electronic states were also shifted to better match the experimental PE transition energies obtained from Ref.~\citenum{bally1998electronic} for Np$^+$, from Refs.~\citenum{boschi1972photoelectron} and~\citenum{szczepanski1993electronic} for An$^+$, and from Ref.~\citenum{boschi1972photoelectron-bis} for Py$^+$. In the following, these shifts are used to identify the spectral positions of the various electronic states, in particular when computing the RF rate constants.\par
\begin{figure}
    \centering
    \includegraphics[width=8.5cm]{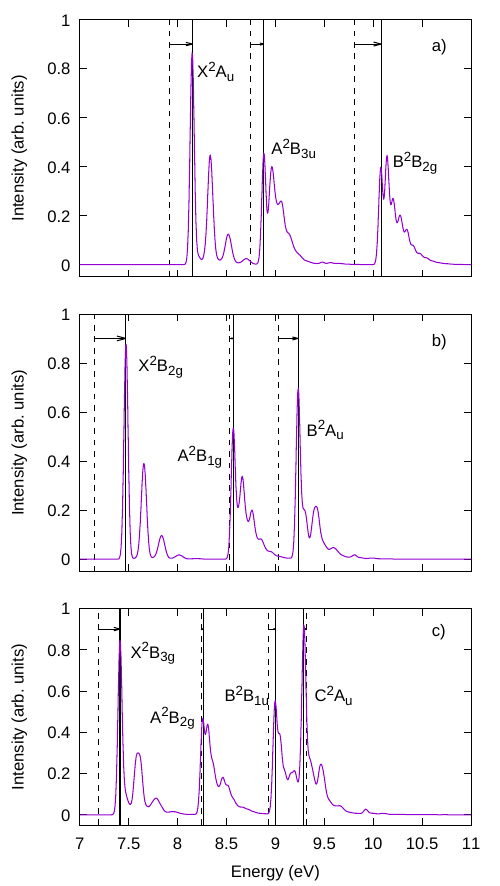}
    \caption{Simulated photoelectron spectra for (a) Np$^+$; (b) An$^+$; and (c) Py$^+$. The spectra are shifted horizontally to match experimental lines. The original positions of the states, calculated by our model, are shown by dashed vertical lines, while the full lines show the shifted positions. All spectra are convoluted with a Gaussian with FWHM of 50~\cm.
    \label{fig:PES}}
\end{figure}
Fig.~\ref{fig:PES}a) shows the harmonic theoretical PE spectra for Np$^+$, including the contributions from states D$_0$, D$_1$ and D$_2$ and assuming that the system initially resides in the vibrational ground state of the neutral molecule. The theoretical spectrum of the S$_0$ $\rightarrow$ D$_0$ transition shows a clear vibrational progression with an origin band located at 8.15~eV and additional bands at 8.33 eV and 8.52 eV, corresponding to a vibrational frequency of 1480~\cm. This value is comparable to the experimental spectrum from Mayer and coworkers,\cite{mayer2011threshold} where a similar vibrational progression with frequency of 1408~\cm\ was reported. Here it should be emphasized that no frequency scaling factor was applied to account for anharmonicities or potential errors inherent to quantum chemical calculations in our calculations. In the case of the S$_0$ $\rightarrow$ D$_1$ and S$_0$ $\rightarrow$ D$_2$ transitions, the PE spectrum of Np$^+$ is broader and without a clear vibrational progression, also in agreement with the measurements.\cite{bally1998electronic} 

Figs. \ref{fig:PES}b) and \ref{fig:PES}c) show the corresponding PE spectra for An$^+$ and Py$^+$, respectively. Similarly as for Np$^+$, clear vibrational progressions are also seen in the S$_0$ $\rightarrow$ D$_0$ transition of both An$^+$ and Py$^+$ cations. In the anthracene case, the bands found at 7.47 eV, 7.65 eV, and 7.84 eV correspond to a vibrational frequency of about 1500~\cm, comparable with the experimental PE spectra of Boschi and coworkers, within 100~\cm\ residual error.\cite{boschi1972photoelectron} In the case of Py$^+$, the bands located at 7.41 eV, 7.60 eV, and 7.78 eV correspond to a vibrational frequency of 1490~\cm, in satisfactory agreement with the values of 1390~\cm\ measured by Boschi and Schmidt\cite{boschi1972photoelectron-bis} and 1389~\cm\ measured by Mayer and cowokers.\cite{mayer2011threshold} Similarly as for Np$^+$, the cationic excited state bands in the PE spectra of the two other cations are broader compared to the S$_0$ $\rightarrow$ D$_0$ transition, without a clear progression and in agreement with PE experiments.\cite{boschi1972photoelectron,szczepanski1993electronic,boschi1972photoelectron-bis,mayer2011threshold}

Besides energetic properties, Table~\ref{tab:electronicstates} also provides the FC and the HT contributions to the oscillator strength involved in Eqs.~(\ref{eq:fFC}) and (\ref{eq:fHT}), between the electronic ground and excited states of the cations. The various oscillator strengths were determined using the theoretical adiabatic transition energies. The HT contribution was calculated at 0~K, and at a more realistic effective temperature of 2000~K that corresponds to about 5~eV lying in the electronic ground state of Np$^+$ (vide infra). Our calculated FC oscillator strengths are of similar magnitude compared to previously published work, although consistently higher. For example, in the case of the D$_0$ $\rightarrow$ D$_2$ transition of Np$^+$, our calculated value of 0.082 is larger than the value of 0.047 reported by Malloci and coworkers\cite{Malloci:2004aa} based on DFT/B3LYP/4–31G, as well as the value of 0.057 given by Hirata and coworkers\cite{Hirata:1999aa} based on DFT/BLYP/6-31G** together with the Tamm–Dancoff approximation. However, our total oscillator strength (FC plus HT) for the D$_0$ $\rightarrow$ D$_2$ transition of Np$^+$ at zero effective temperature compares well with the value of 0.09 given by Lee and coworkers based on the DFT/LC-$\omega$HPBE/cc-pVTZ method.\cite{Lee:2023aa}

\begin{figure}
    \centering
    \includegraphics[width=8.5cm]{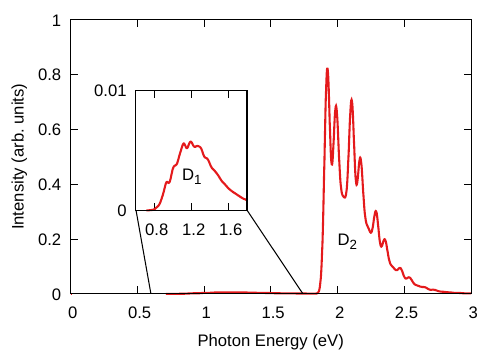}
    \caption{Simulated absorption spectrum for Np$^+$ at zero temperature, convoluted using a Gaussian function with FWHM of 400~\cm. The inset highlights the 0.6--1.8~eV energy range.}
    \label{fig:ABS}
\end{figure}

Except for symmetry-forbidden transitions, the FC contribution is found to be dominant compared to the HT contribution for all transitions and all cations considered. However, the HT contribution becomes non negligible in some cases, especially for Py$^+$ where it can amount to more than a fifth of the total oscillator strength. For all transitions, the HT contribution increases with increasing effective temperature. Such an increase varies in magnitude depending on the cation and transition, and can be relatively minor (18\% for the D$_0$ $\rightarrow$ D$_3$ transition of Py$^+$) or conversely quite strong (166\% for the D$_0$ $\rightarrow$ D$_2$ transition of Np$^+$). These variations arise from differences in the transition dipole moment surface. Specifically, transition where the transition dipole moment varies the most with low-frequency vibrational modes are found to exhibit the largest increase as a function of the effective temperature. For the D$_0$ $\rightarrow$ D$_1$ transitions, which are forbidden by symmetry, the HT contribution to the oscillator strength does not vanish but remains small compared to the FC oscillator strength of symmetry-allowed transitions. For instance, in the case of Np$^+$, the total oscillator strength for the D$_0$ $\rightarrow$  D$_1$ transition is about two orders of magnitude weaker than that of the D$_0$ $\rightarrow$  D$_2$ transition. Such a difference is illustrated in Fig.~\ref{fig:ABS}, showing the simulated absorption spectrum for Np$^+$ obtained using Eq.~(\ref{eq:absspectrum}) and assuming a 0~K equilibrium temperature, the individual peaks being convoluted with a 400~\cm\ Gaussian function.
%

The theoretical absorption spectrum displays a strong band corresponding to the D$_0$ $\rightarrow$ D$_2$ transition, with a clear vibrational progression also seen in the experimental spectrum.\cite{bally1998electronic,Pino:1999bh} To better appreciate the overall quality of our modeling, Fig.~\ref{fig:abspino} in appendix~\ref{sec:app_absorption} provides a detailed comparison of our computed absorption spectrum with the experimental data of Pino and coworkers (Ref.~\citenum{Pino:1999bh}). In our computed spectrum, a very weak absorption band is also found, corresponding to the D$_0$ $\rightarrow$ D$_1$ transition, with an intensity more than two orders of magnitude weaker than the D$_0$ $\rightarrow$ D$_2$ band. This ratio is in accordance with the ratio between the transition dipole moments of the two states, as given in Table~\ref{tab:electronicstates}. Such a weak signal is also consistent with experimental measurements,\cite{bally1998electronic} in which no contribution from the D$_0$ $\rightarrow$ D$_1$ transition was reported, the D$_0$ $\rightarrow$  D$_1$ contribution being thus likely hidden in the noise.

\begin{figure}
    \centering
    \includegraphics[width=8.5cm]{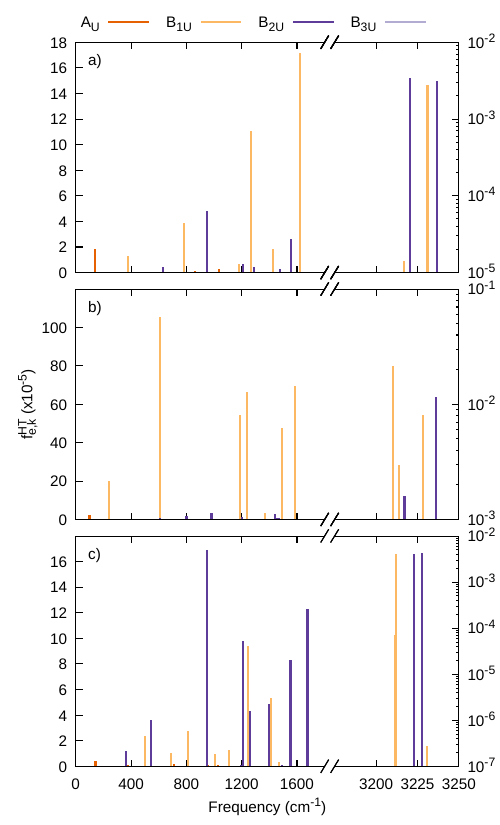}
    \caption{Mode-specific Herzberg-Teller oscillator strength $f_{e,k}^{\text{HT}}(T)$ for the D$_0$ $\rightarrow$  D$_1$ transition of the cations (a) Np$^+$; (b) An$^+$; and (c) Py$^+$, at $T=2000$~K and as a function of the harmonic frequency $\omega_{e,k}$. The gerade vibrational modes have no contribution. The oscillator strengths in the CH stretching region are given in logarithmic scale for better visualization.}
    \label{fig:modes}
\end{figure}
The overall HT oscillator strength can be decomposed over the various normal modes. This decomposition is illustrated in Fig.~\ref{fig:modes}, where the HT oscillator strength $f^{\text{HT}}_{e,k}$  from each normal mode for the symmetry-forbidden transition D$_0$ $\rightarrow$ D$_1$ of Np$^{+}$, An$^+$ and Py$^+$  at $T=2000$~K is obtained from Eq.~(\ref{eq:fHT}). For all molecules, as expected, only the ungerade normal modes that break the inversion symmetry produce a transition dipole moment. Fig.~\ref{fig:modes} shows that the high-frequency CH stretching region only gives a negligible contribution to the transition dipole moments and that the normal modes contributing the most to the oscillator strength are of B$_{1u}$ and B$_{2u}$ symmetries.
In particular, for Np$^+$, the two modes having the highest contribution are both of B$_{1u}$ symmetry with frequencies $1618.9~\cm$ and $1266.2~\cm$. These two modes correspond to combinations of CC stretching and CH in-plane bending motions. In the case of An$^+$, the strongest contributing mode is a B$_{1u}$ mode with frequency $607.7~\cm$ corresponding mostly to collective CC stretching motion. For Py$^+$, the main contribution arises from a B$_{2u}$ mode with frequency  $946.0~\cm$ also corresponding to a collective CC stretching motion. For the three cations, a small contribution from a completely antisymmetric A$_u$ vibrational mode is also noticed in the very low frequency range ($139.7~\cm$, $99.6~\cm$ and $143.3~\cm$ for Np$^{+}$, An$^+$ and Py$^+$, respectively). 

Our model, which quantifies the effect of molecular deformation on the electronic transition dipole moment, relies on the linear expansion of the transition dipole moment [Eqs.~(\ref{eq:transdipoleground}) and~(\ref{eq:transdipoleexcited})] and the harmonic approximation. Appendix~\ref{sec:harmonic} shows that both assumptions are well satisfied for the present systems under the operating conditions of internal energies lower than 13.6~eV.

\subsection{Effective canonical temperature}
\label{sec:Teff}
\begin{figure}
    \centering
    \includegraphics[width=8.5cm]{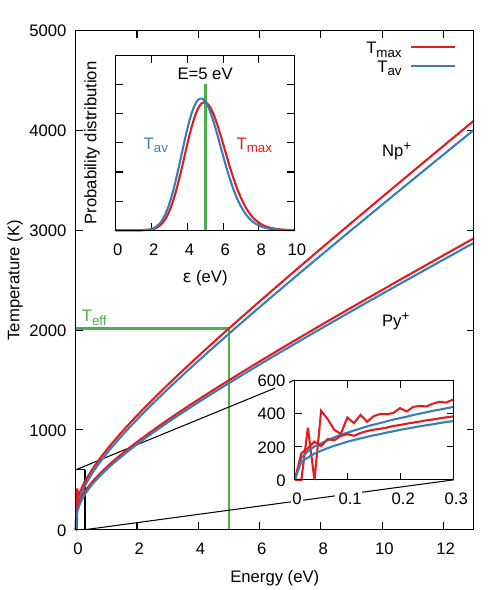}
    \caption{Average temperature $T_{\text{av}}$ and maximum temperature $T_{\text{max}}=T_{\mu}$ as a function of the microcanonical energy for the D$_0$ states of Np$^+$ and Pyr$^+$. The inset at the upper left shows the canonical probability distributions of internal energies for the D$_0$ state of Np$^+$ obtained with $T_{\text{av}}$ and $T_{\text{max}}$ as effective temperatures at 5~eV microcanonical energy. The inset at the lower right highlights the discontinuous variations at low energy.}
    \label{fig:comp_distrib}
\end{figure}
In our model, Eq.~(\ref{eq:rfrate3}) is solved by replacing the original microcanonical statistical distribution by an equivalent canonical distribution at an effective temperature $T_{\rm eff}$. The equivalence between the microcanonical and canonical ensembles strictly holds only in the thermodynamic limit, for example in very large systems. However, for finite-size systems, in particular for small molecules with few degrees of freedom such as Np$^+$, this equivalence may break down. In this study, we show that this assumption remains valid even for Np$^+$, the smallest PAH considered here.  In addition, we explore two possible choices for $T_{\rm eff}$ in this approximation  based on adjusting either the maximum ($T_{\rm max}$) or the average ($T_{\rm av})$ of the canonical distribution to match the microcanonical energy $E_0$. 

These choices are illustrated in the upper left inset of Fig.~\ref{fig:comp_distrib}, where the canonical probability distributions in the D$_0$ state of Np$^+$ evaluated with the two temperatures $T_{\rm max}$and $T_{\rm av}$ and corresponding to 5~eV microcanonical energy, are shown as a function of internal energy. For this harmonic system, and for any electronic state $i$, the average energy $\langle E\rangle$ from which the effective temperature $T_{\rm av}$ is obtained is an explicit function of the normal mode frequencies,
\begin{equation}
\langle E \rangle (T) = \sum_{k=1}^n  \frac{\hbar\omega_{i,k}}{e^{\beta\hbar\omega_{i,k}}-1}.
\end{equation}

For $T_{\text{max}}$, the peak of the probability distribution is, by definition, precisely centered at the microcanonical energy, while for $T_{\text{av}}$ the corresponding peak is slightly shifted towards lower energies. This is a direct result of the asymmetry of the Boltzmann distribution. Fig.~\ref{fig:comp_distrib} also shows the energy versus temperature variations for both the  microcanonical ($T_{\rm max}=T_\mu$) and the canonical ($T_{\rm av}$) temperatures in the case of the D$_0$ state of Np$^+$ and in the case of the D$_0$ state of Py$^+$. As expected for an harmonic system, both temperatures increase linearly at high energies, but with slightly different slopes. As a result, the canonical temperature is somehow slightly lower than the microcanonical temperature. The difference between the two temperatures is directly related to the finite size of the system, and illustrates that the two statistical ensembles are equivalent only in the thermodynamic limit. This can be seen in the main Fig.~\ref{fig:comp_distrib} which compares the two choices of effective canonical temperatures for Np$^+$ and Py$^+$. In Fig.~\ref{fig:comp_distrib}, the canonical temperature is a well-defined, monotonic and therefore bijective function of energy. However, such is not the case for the microcanonical temperature which is only defined for energies that are large enough for the density of states to be approximated by a continuous and differentiable function. This can be seen as a cusp appearing at low energy in the microcanonical temperature represented in Fig.~\ref{fig:comp_distrib}, and highlighted as an inset in the lower right part of this figure.

\begin{figure}
    \centering
    \includegraphics[width=8.5cm]{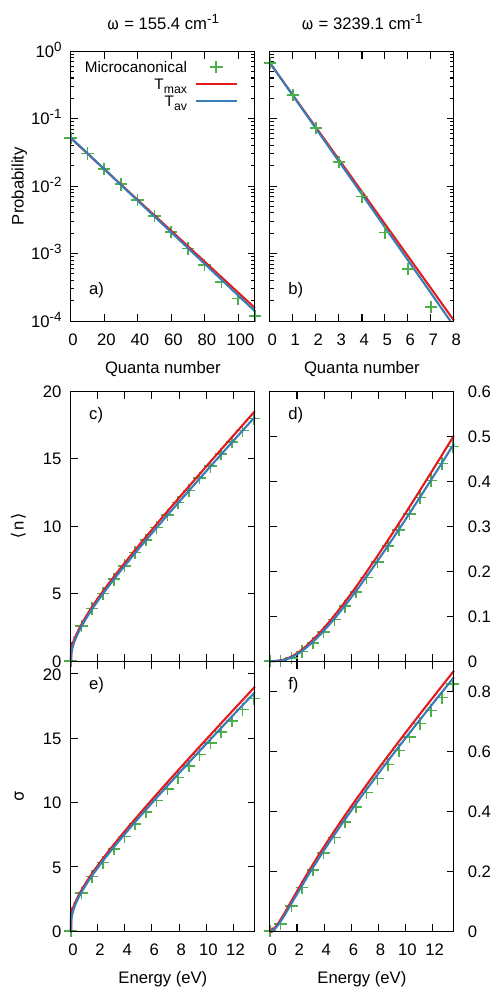}
    \caption{Statistics of the naphthalene cation in D$_0$ state being in specific vibrational modes as a function of internal energy, from the true microcanonical ensemble, compared with the predictions in the canonical ensemble with effective temperatures taken as $T_{\rm max}$ or $T_{\rm av}$, as defined in the text. (a,b) Probability $P_m^k$ to have $m$ quanta in mode $k$ at internal energy $E_0=13.6$ eV ; (c,d) average numbers of quanta; (e,f) mean-square fluctuation around the average numbers. (a,c,e) Lowest-frequency mode with $\omega=155.4$~\cm; (b,d,f) Highest-frequency mode with $\omega=3239.1$~\cm.
    }
    \label{fig:temperatures}
\end{figure}
In addition to the effective temperature, the probability $P^k_m(E_0)$ to find $m$ quanta in mode $k$ at internal energy $E_0$ was determined in the original microcanonical ensemble, as well as in the canonical ensembles using either $T_{\rm max}$ or $T_{\rm av}$ as the effective canonical temperature. In the microcanonical ensemble, and for any electronic state $i$, the probability $P^k_m(E_0)$ is defined by
\begin{equation}
P^k_m(E_0) = \frac{\Omega^{(k)}_i(E_0-m\hbar\omega_{i,k})}{\Omega_i(E_0)},
\label{eq:microprob}
\end{equation}
where $\Omega^{(k)}_i$ is the vibrational density of states excluding the vibrational mode $k$. In the canonical ensemble, this probability is given by
\begin{equation}
P^k_m(T_{\text{eff}}) = e^{-m\beta_{\text{eff}}\hbar\omega_{i,k}}\left(1-e^{-\beta_{\text{eff}}\hbar\omega_{i,k}}\right).
\label{eq:canprob}
\end{equation}
The variations of this quantity with increasing number of quanta are shown in Figs.~\ref{fig:temperatures}a) and~\ref{fig:temperatures}b) for the lowest-frequency ($\omega_{g,1}=155.4~\cm$) and the highest-frequency ($\omega_{g,n}=3239.1~\cm$) modes of the D$_0$ state of Np$^+$ and at the highest energy considered in this work, $E_0=13.6$~eV. As expected, this probability decreases exponentially when described in the canonical ensemble, whatever the underlying choice for the corresponding temperature between $T_{\rm max}$ and $T_{\rm av}$. In contrast, the true microcanonical probability decreases exponentially for small values of $n$ but deviates from this behavior at larger $n$. Overall, the differences between the various probabilities are small but the canonical temperature value $T_{\rm av}$ that originates from matching $E_0$ to $\langle E\rangle$ provides the best agreement to the microcanonical reference. This happens to hold for all values of the internal energy and for all vibrational modes, as evaluated from the statistics in the number of quanta $n_k$ as a function of internal energy $E$. Figs.~\ref{fig:temperatures}c) and \ref{fig:temperatures}d) show the variations of the average numbers $\langle n_k\rangle$ for the lowest- and highest-frequency modes of Np$^+$ at 155.4~\cm\ and 3239.1~\cm, respectively, while Figs.~\ref{fig:temperatures}e) and \ref{fig:temperatures}f) display the mean square fluctuations $\sigma_k(E)$ associated with the corresponding microcanonical [Eq.~(\ref{eq:microprob})] and canonical [Eq.~(\ref{eq:canprob})] distributions.

For both modes, the average number of quanta calculated in the canonical ensemble employing temperature $T_{\rm av}$ is almost identical to the reference value obtained in the microcanonical ensemble. This result was expected since the canonical temperature is defined through the average number of quanta. By looking at the deviation, residual differences between the distributions appear but the canonical temperature $T_{\rm av}$ is consistently a better choice compared to the alternative choice $T_{\rm max}=T_\mu$ of the true microcanonical temperature. Such conclusions also hold for the other cationic PAH and different electronic states considered in this work (results not shown). This justifies our choice of the canonical ensemble with effective temperature $T_{\rm eff}=T_{\rm av}$ when modeling the RF differential rate constants.

\subsection{RF rate constants}
\begin{figure}
    \centering
    \includegraphics[width=8.5cm]{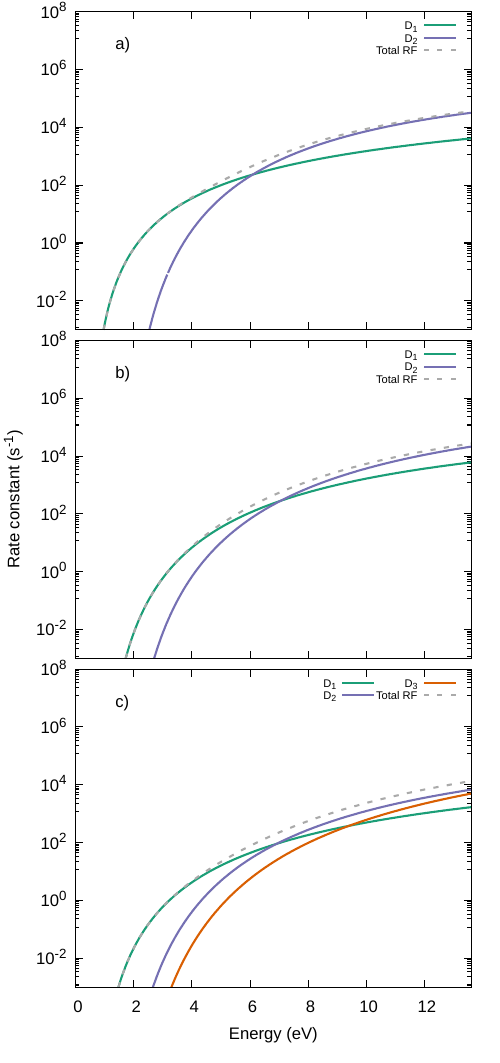}
    \caption{State-resolved (solid lines) and cumulated (dashed lines) recurrent fluorescence rate constants for (a) Np$^+$, (b) An$^+$, and (c) Py$^+$, as a function of internal energy.}
    \label{fig:rc}
\end{figure}
Having validated the canonical framework for computing the RF rate constants through Eq.~(\ref{eq:rfratecstfinal}) with the temperature $T_{\rm av}$ serving as the effective canonical temperature, Fig.~\ref{fig:rc} shows the variations of the total rate constants [see Eq.~(\ref{eq:totalRF})] for the three cations as a function of the internal energy, highlighting the various contributions from individual electronic states. The positions of the electronic excited states were shifted  to better match experimental data with the same shift used for the PE spectra. These shifts are essential in determining the occupation probabilities of the various electronic excited states. It is worth noting that, for Np$^+$, the D$_0$ $\rightarrow$ D$_2$ transition is experimentally located at 1.93 eV,\cite{bally1998electronic} which is slightly higher than the value of 1.85 eV reported by experimental {\em absorption} spectroscopy.\cite{Pino:1999bh} The residual 650 cm$^{-1}$ discrepancy may be attributed to the lower resolution of the PE spectra in these experiments. Nevertheless, since absorption spectroscopy does not provide information on the position of the D$_1$ state, the  experimental value associated with the PE spectrum\cite{bally1998electronic} was chosen here to ensure a consistent approach.

In this study, the RF rate constants are computed across a broad range of internal energies, up to 13.6 eV. It is important to note that, for Np$^+$, IR fluorescence dominates at low energies (typically below 5 eV), while unimolecular dissociation becomes the primary process at higher energies (above 6.5 eV). Our focus, however, is not on simulating the entire relaxation cascade. Instead, we investigate how the presence of multiple electronic excited states influences both the efficiency and spectral profile of RF emission, an effect that grows more pronounced with increasing molecular size, especially as dissociation is expected to become less and less efficient.

While the symmetry-forbidden D$_0$ $\rightarrow$ D$_1$ transition contributes negligibly compared to symmetry-allowed transition in the absorption spectrum, it can make a comparable or even dominant contribution to the RF emission spectrum depending on internal energy. For Np$^+$, at energies below the threshold of the D$_2$ state (1.93 eV), the D$_1$ $\rightarrow$ D$_0$ transition is obviously dominant, D$_1$ being the only state that can contribute. However, above the D$_2$ state threshold, the D$_1$ $\rightarrow$ D$_0$ transition remains dominant until the internal energy reaches $E = 6.41$~eV. Despite its very low oscillator strength, the significant contribution of the D$_1$ $\rightarrow$ D$_0$ transition arises from the much lower adiabatic energy of the D$_1$ state compared to the D$_2$ state, which leads to a stronger population in the D$_1$ state and compensates for the weak transition dipole moment. At higher energies, the D$_2$ $\rightarrow$ D$_0$ transition eventually dominates the total RF rate constant, contributing up to 7.45 times more than the D$_1$$\rightarrow$ D$_0$ transition at $E = 13$~eV.

In the case of the An$^+$ and Pyr$^+$ cations, similar trends arise. For An$^+$, the D$_1$ $\rightarrow$ D$_0$ transition prevails over the D$_2$ $\rightarrow$ D$_0$ transition up to $E = 7.04$~eV and, at $E = 13$ eV, its contribution is only 3.33 times weaker than that of the D$_2$ $\rightarrow$ D$_0$ transition. For Pyr$^+$, the emission from the D$_1$ state exceeds the emission from the D$_2$ state for energies below $E = 6.91$~eV, and the emission from the D$_3$ state for energies below $E = 9.39$ eV.

\begin{figure*}
    \centering
    \includegraphics[width=17cm]{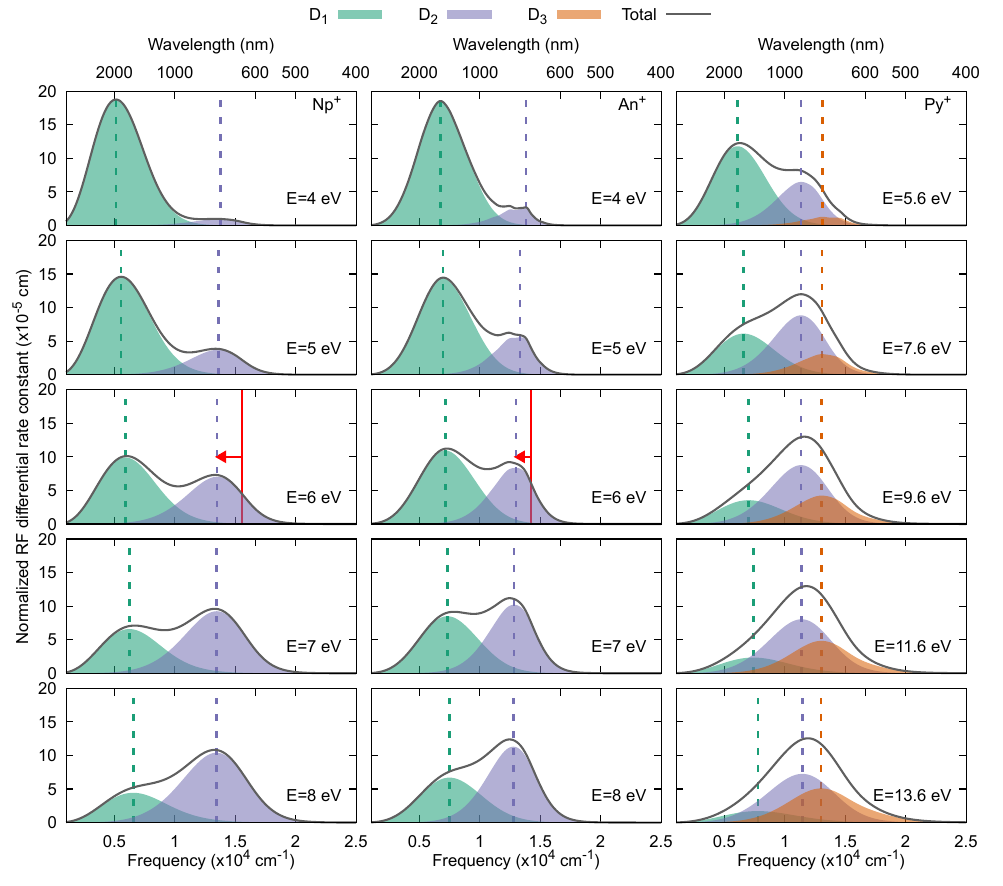}
    \caption{Normalized RF differential rate constants for Np$^+$ (left panels), An$^{+}$ (middle panels) and Py$^+$ (right panels) as a function of the photon frequency $\omega$ and for various internal energies, highlighting the contributions from the D$_1$, D$_2$ and D$_3$ electronic states. The dashed vertical lines indicate the positions of the maxima in the differential rate constants for the D$_1$ $\rightarrow$ D$_0$ and  D$_2$ $\rightarrow$ D$_0$ transitions. For Np$^+$ and An$^+$, solid red vertical lines mark the positions of the D$_0$ $\rightarrow$ D$_2$ absorption transitions, illustrating the Stokes shift at 6 eV internal energy.}
    \label{fig:RCfreq}
\end{figure*}
In Fig.~\ref{fig:RCfreq}, the normalized differential rate constants for Np$^+$,  $k^{\text{RF}}(\omega;E)/k^{\text{RF}}_{\text{tot}}(E)$, are represented as a function of the emitted photon frequency and for internal energies in the 4--8 eV range. The contributions from both electronic excited states are shown separately, highlighting the decreasing importance of the D$_1$ state, relative to D$_2$, as energy is increased. This is also in accordance with the conclusions from Fig.~\ref{fig:rc}. Across all displayed energies, the differential rate constant shows a double peak behavior that emphasizes the separate contributions from the two electronic transitions. For each band, the emission spectrum is broad and lacks vibrational structure, which is directly attributable to the high excess energy in the molecule, resulting into a large number of vibrational quanta distributed over many modes. The linewidths of both the D$_1$ and D$_2$ bands increase with increasing energy, also due to the larger number of vibrational quanta. The D$_1$ band exhibits a FWHM of 4860~\cm\ at $E = 4$ eV and 6220~\cm\ at $E =8$ eV. For the D$_2$ band, the FWHM is 4520~\cm\ at $E = 4$ eV and 6280~\cm\ at $E = 8$ eV. While the position of the maximum of the D$_2$ band remains nearly constant around $\omega=13600~\cm$, the maximum of the D$_1$ band shifts significantly from 5120~\cm\ to 6560~\cm. This behavior originates from the $\omega^3$ term in the expression of the rate constant [see Eq.~(\ref{eq:rfratecstfinal})], which favors higher frequencies. Since the D$_1$ band is relatively low in energy, it is particularly sensitive to this term, causing the peak of the differential rate constant to shift toward higher frequencies. In contrast, the D$_2$ band, being higher in energy, is only barely affected by this effect.

The variations of the normalized differential rate constants for An$^+$ are shown in Fig.~\ref{fig:RCfreq} as well. As is the case for Np$^+$, the contributions from the two bands D$_1$ $\rightarrow$ D$_0$ and D$_2$ $\rightarrow$ D$_0$ are also clearly seen. Because the D$_1$ $\rightarrow$ D$_0$ transition is located at a higher energy, it only shifts from $\omega=6760~\cm$ at 4 eV to 7500~\cm\ at 8 eV. At the low energy $E=4$ eV, the D$_2$ band exhibits a small vibrational structure with two peaks located at $\omega=12560~\cm$ and 13820~\cm. This vibrational feature disappears at higher energies and the band is located at $\omega=12980~\cm$ at $E=8$ eV.

The corresponding rate constants for the pyrene cation, also represented in Fig~\ref{fig:RCfreq}, now highlight the contribution from the D$_3$ state, in addition to the lower D$_1$ and D$_2$ states. At low energy, the total emission spectrum exhibits a double-peak structure corresponding to the bands  D$_2$ $\rightarrow$ D$_0$ and  D$_1$ $\rightarrow$ D$_0$. As the energy increases, the energy of the D$_1$ $\rightarrow$ D$_0$ transition located near 7000~\cm\ weakens while the D$_2$ $\rightarrow$ D$_0$ band located at $\omega=11400~\cm$ becomes stronger. At very high energy, the contribution of the D$_3$ $\rightarrow$ D$_0$ transition located around $13200~\cm$ equals the contribution of the D$_2$ $\rightarrow$ D$_0$ band, producing a single broad peak around $12000~\cm$.

\subsection{Experimental comparison}

Saito and coworkers measured the RF emission spectrum of Np$^+$ using various bandpass filters.\cite{saito2020direct} In this experimental work, the cations were excited at an average internal energy of approximately 6.5~eV. The spectrum recorded by these authors initially increases around 560~nm ($\sim 17900~\cm$) and reaches a maximum between 750 and 800 nm ($\sim$ 13330--12500~\cm). Comparing with the absorption spectra experimentally measured for Np$^+$ by Bally and coworkers,\cite{bally1998electronic} and independently by Pino and coworkers,\cite{Pino:1999bh} the RF emission spectrum reported by Saito and coworkers agrees in the increasing region between 560 nm up to the maximum of the absorption spectrum at 670.9 nm (14900~\cm),\cite{Pino:1999bh} but shows a marked difference at higher wavelengths where the absorption spectrum vanishes. More recently, Kusuda and coworkers\cite{kusuda2024detection} measured the RF emission spectrum of An$^+$ with a similar experiment as for Np$^+$. The average internal energy in the cations was around 6~eV. In Ref.~\citenum{kusuda2024detection}, the measured RF emission spectrum steadily increases from 625~nm and reaches a maximum at 775~nm, before remaining almost constant until the measured wavelength reaches its maximum at 825~nm. Similarly as for the Np$^+$ experiment,\cite{saito2020direct} the position of the maximum of the RF emission spectrum (775~nm, 12900~\cm) is observed at a lower frequency than the maximum of experimental absorption spectrum located at 722.4~nm (13843~\cm) and measured in Ref.~\citenum{szczepanski1993electronic}. Both shifts measured by Saito and coworkers for Np$^+$ and An$^+$  were interpreted in Ref.~\citenum{saito2020direct} and~\citenum{kusuda2024detection} as Stokes shifts. 

As shown in Fig.~\ref{fig:RCfreq}, in our model the absorption peak for the transition D$_0$ $\rightarrow$ D$_2$ of Np$^+$ is predicted at 15565~\cm\ (or 1.93~eV, 642~nm). The difference between this theoretical value and the experimental absorption peak position is directly due to our calibrating  the electronic states based on experimental PE data. The peak of the RF emission spectra is predicted to be located at 13600~\cm\ (735~nm) with a value that is nearly independent of the internal energy. The Stokes shift predicted with our model  between absorption and emission, highlighted in Fig.~\ref{fig:RCfreq},  is around 2000~\cm, a value in line with the shift observed experimentally by Saito and coworkers ($\sim$1500--2500~\cm).\cite{saito2020direct}

In the case of An$^+$, the peak position for the D$_0$ $\rightarrow$ D$_2$  transition (see Fig.~\ref{fig:RCfreq}) is predicted by our model to be located at 14195~\cm\ (1.76 eV, 705 nm) a value relatively close to the experimental absorption peak located at 13843~\cm\ (722 nm). Our emission model predicts a peak fluorescence around 12980~\cm\ corresponding to a predicted Stokes shift of 1200~\cm. This value is slightly larger than the value observed by Kusuda and coworkers (around 1000~\cm).\cite{kusuda2024detection}

As shown in Fig.~\ref{fig:RCfreq}, the position of the D$_1$ $\rightarrow$ D$_0$ transition for all molecules is too low to be resolved with the bandpass filters used in Refs.~\citenum{saito2020direct} and ~\citenum{kusuda2024detection}. However, depending on the internal energy, our model predicts a very large contribution of this transition to the total RF cooling rate. This transition should be clearly observable in the wavelength range of 1000~nm to 2000~nm.

\section{Conclusions and perspectives}
\label{sec:conclusions}
Recurrent fluorescence is a crucial relaxation mechanism that enhances the stability of polycyclic aromatic hydrocarbons and other molecular systems in the interstellar medium. In this work, a novel statistical model was introduced to describe RF rate constants, explicitly incorporating Herzberg–Teller and Duschinsky rotation effects, as well as a full account of vibrational progressions within the harmonic approximation. Being aimed at isolated molecules, the model is intrinsically built on microcanonical statistics. However, a canonical framework was found to be more effective and even quantitative, provided that the temperature is chosen so the microcanonical energy matches the internal energy in the canonical ensemble.

Application of the model requires ingredients that can be obtained from quantum chemistry calculations, here based on density-functional theory in its static and time-dependent versions for the highly symmetric naphthalene, anthracene, and pyrene cations. Our study confirms the observation of Saito and coworkers\cite{saito2020direct,kusuda2024detection} that the increased emission in the long wavelength compared to absorption can be attributed to broadening and a Stokes shift of the D$_2$ $\rightarrow$ D$_0$ band.  Our study also suggests that, for these systems, the low-lying, symmetry-forbidden electronic transitions may contribute more significantly to their cooling efficiency than higher-energy, symmetry-allowed transitions, particularly for energies typically below 6.5~eV. This energy range is of particular interest because, above this threshold, fragmentation likely dominates the relaxation cascade, whereas below approximately 5~eV, infrared photon emission is the primary relaxation pathway.\cite{Lee:2023aa} This range of interest precisely corresponds to the typical energies probed in experiments using ion traps or storage rings, such as those conducted by Saito and coworkers\cite{saito2020direct} on the naphthalene and anthracene cations. Therefore, our findings not only advance the understanding of PAH relaxation kinetics in laboratory settings but also emphasize the critical role of forbidden transitions in PAH survival under actual astrochemical environments.

Future efforts will be aimed at extending the model to quantify the impact of low-lying forbidden transitions on the relaxation cascade of various PAHs. Such extensions require explicitly accounting for competing relaxation processes, including infrared emission, fragmentation, as well as isomerization. Because some of these processes are intrinsically anharmonic, it would be useful to extend the present model beyond the harmonic approximation, either through perturbative expansions or molecular dynamics simulations, ideally corrected for nuclear quantum effects.

At present, the model has been solely used to describe recurrent fluorescence from an electronic excited state back to the electronic ground state. In a straightforward extension, the contribution of recurrent fluorescence between two electronic excited states could be also considered. Additionally, the contribution of inverse fluorescence, a process involving fluorescence from a vibrationally excited electronic ground state to an electronic excited state, could be quantified as well. Despite being mentioned as a possible mechanism by Léger and coworkers several decades ago,\cite{leger1988predicted} this contribution, though expectedly small, remains to be quantitatively assessed.

\appendix
%

%
%
%
%

\section{Duschinsky parameters}
\label{sec:app_duschinsky}

The differences between the potential energy surfaces in the electronic ground and excited states cause variations in the equilibrium position and the normal modes, as expressed by Eq.~(\ref{eq:duschinsky}).
For a $N$-atom system, and denoting by $\vect{x}=(x_1,\dots,x_{3N})$ the set of $3N$ molecular mass-scaled rectilinear coordinates, the electronic ground state potential energy surface is expanded around its equilibrium position $\vect{x}_g$ as
\begin{equation}
V_g(\vect{x}) \approx \frac{1}{2} \left(\vect{x}^T-\vect{x}_g^T\right) \vect{V}_g \left( \vect{x}-\vect{x}_g\right),
\end{equation}
where $\vect{V}_g$ is the force constant matrix. We further denote by $\vect{L}_g$ the $3N\times n$ matrix composed of the $n=3N-6$ vibrational eigenstates of $\vect{V}_g$, and write the ground state vibrational normal modes as
\begin{equation}
\vect{x} = \vect{x}_g + \vect{L}_g \vect{q}_g.
\end{equation}
A similar harmonic expansion can be performed in the electronic excited state,
\begin{equation}
V_e(\vect{x}) \approx \frac{1}{2} \left(\vect{x}^T-\vect{x}_e^T\right) \vect{V}_e \left( \vect{x}-\vect{x}_e\right) + \hbar\omega_{eg},
\end{equation}
where the equilibrium coordinates $\vect{x}_e$ are expressed in the Eckart frame with respect to the electronic ground state equilibrium geometry $\vect{x}_g$. Similarly as with the electronic ground state, the $3N\times n$ matrix $\vect{L}_e$ is introduced, combining the $n$ vibrational eigenstates of $\vect{V}_e$, allowing the electronic excited state vibrational normal modes to be defined as
\begin{equation}
\vect{x} = \vect{x}_e + \vect{L}_e \vect{q}_e.
\end{equation}
Neglecting dynamical axis-switching effects,\cite{Lucas:1973aa,Ozkan:1990aa} the Duschinsky relation connecting $\vect{q}_g$ and $\vect{q}_e$ is linear and written as
\begin{equation}
\vect{q}_e = \vect{L}_e^T \vect{L}_g \vect{q}_g + \vect{L}^T_e \left(\vect{x}_g-\vect{x}_e\right),
\end{equation}
from which the expression for the Duschinsky rotation matrix $\vect{J}$ and the displacement vector $\vect{K}$ can be readily obtained as
\begin{equation}
\vect{J} = \vect{L}_e^T \vect{L}_g,
\end{equation}
and
\begin{equation}
\vect{K} =  \vect{L}^T_e \left(\vect{x}_g-\vect{x}_e\right),
\end{equation}
respectively.

\begin{figure}
    \centering
    \includegraphics[width=8.5cm]{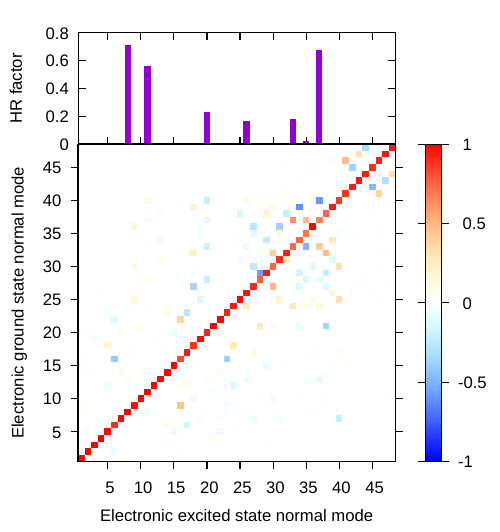}
    \caption{Magnitude of the eigenmodes on the D$_1$ electronic state, as a function of the reference eigenmode on the electronic ground state, for the Np$^+$ cation. The upper graph shows the corresponding Huang-Rhys (HR) factors that measure the magnitude of the electronic-vibronic coupling.}
    \label{fig:duschmat}
\end{figure}

For the present cationic PAHs, the Duschinsky matrices are almost entirely diagonal, due to the similarity between the electronic ground and relevant excited state surfaces. This is illustrated in Fig.~\ref{fig:duschmat}, which shows the vibrational eigenmodes of the electronic ground state of the Np$^+$ cation, projected onto the eigenmodes of the D$_1$ electronic excited state, the various modes being rearranged so their projection is maximum in magnitude, for any given mode on the reference electronic excited state.
The magnitude of the deformation between the electronic ground and excited states equilibrium positions was quantified  using the dimensionless Huang-Rhys factors\cite{Huang:1950kl} defined here by
\begin{equation}
S_{k} = \frac{1}{2\hbar}\omega_{e,k}K_{k}^2 ,
\end{equation}
and superimposed as an histogram in Fig.~\ref{fig:duschmat}. Consistently with the Duschinsky matrix, only a few modes exhibit significant couplings, and a similar conclusion is reached for the 6 other transitions discussed earlier (results not shown).
\section{Calculation of the dipole moment time autocorrelation function}
\label{sec:app_correl}
Following Ianconescu and Pollak,\cite{Ianconescu:2004aa} the trace in Eq.~(\ref{eq:correl1}) can be solved using the ket vectors in coordinate representation $\ket{\vect{q}_g}$ and $\ket{\vect{q}_e}$. Both sets of vectors are orthonormal and related to each other through
\begin{equation}
\braket{\vect{q}_e}{\vect{q}_g} = \delta\left( \vect{q}_e - \vect{J}\vect{q}_g - \vect{K} \right).
\end{equation}
The correlation function $C_{ge}(\tau_g,\tau_e)$ is formally given by
\begin{equation}
    C_{ge}(\tau_g,\tau_e) = \int \ud^n \vect{q}_g \ud^n \vect{q}'_g  \  \vect{d}(\vect{q}_g)\cdot \vect{d}(\vect{q}'_g)\ 
    \bra{\vect{q}_g}e^{-\frac{\ii}{\hbar}H_g\tau_g}\ket{\vect{q}'_g}
    \bra{\vect{J}\vect{q}'_g+\vect{K}}e^{-\frac{\ii}{\hbar}H_e\tau_e}\ket{\vect{J}\vect{q}_g+\vect{K}},
\end{equation}
and in the harmonic approximation the matrix elements of the propagators can be computed using path integrals,\cite{Zinn-Justin:2005aa} yielding
\begin{equation}
\bra{\vect{q}}e^{-\frac{\ii}{\hbar}H\tau}\ket{\vect{q}'} = \sqrt{\frac{\det \vect{a}}{\left(2\ii \pi \hbar\right)^n}}\exp\left(\frac{\ii}{2\hbar}\left[ \vect{q}^T \vect{b} \vect{q} + \vect{q}^{\prime T} \vect{b} \vect{q}'  - 2\vect{q}^T \vect{a} \vect{q}' \right]\right),
\end{equation}
where the diagonal matrices $\vect{a}$ and $\vect{b}$ are given from their elements $a_{ij}$ and $b_{ij}$ as
\begin{align}
&a_{ij} = \delta_{ij} \frac{\omega_i}{\sin(\omega_i\tau)},\label{eq:amat}  \\
&b_{ij} = \delta_{ij} \frac{\omega_i}{\tan(\omega_i\tau)}.\label{eq:bmat}
\end{align}
Both $\vect{a}$ and $\vect{b}$ are defined for each electronic state $g$ and $e$ and depend on the parameters $\tau_g$ and $\tau_e$. As in Ref.~\citenum{Ianconescu:2004aa}, the following $n\times n$ matrices are next introduced
\begin{align}
    &\vect{A}=\vect{a}_g+\vect{J}^T\vect{a}_e\vect{J},\\
    &\vect{B}=\vect{b}_g+\vect{J}^T\vect{b}_e\vect{J},\\
    &\vect{E}=\vect{b}_e-\vect{a}_e,
\end{align}
providing a simplified expression for the correlation function as
\begin{multline}
C_{ge}(\tau_g,\tau_e)=  \sqrt{\frac{\det \vect{a}_g\det \vect{a}_e}{(2i\pi\hbar)^{2n}}} e^{-i\omega_{eg}\tau_e}\int \ud^n\vect{q}_g \ud^n\vect{q}_g' \ \vect{d}(\vect{q}_g)\cdot \vect{d}(\vect{q}'_g) \\
        \times \exp\left(\frac{\ii}{2\hbar}\left[\vect{q}_g^T \vect{B} \vect{q}_g+\vect{q}_g^{\prime T} \vect{B} \vect{q}'_g-2 \vect{q}_g^T\vect{A}\vect{q}'_g+2\vect{K}^T \vect{E}\vect{J}(\vect{q}_g+\vect{q}'_g)+2\vect{K}^T\vect{E}\vect{K}\right]\right).
\end{multline}
Following Ref.~\citenum{Baiardi:2013aa}, we next introduce the unitary transformation
\begin{align}
    &\vect{u}=\frac{1}{\sqrt{2}}\left(\vect{q}_g+\vect{q}_g'\right),\\
    &\vect{v}=\frac{1}{\sqrt{2}}\left(\vect{q}_g-\vect{q}_g'\right).
\end{align}
Using these variables, the scalar product of dipole moment vectors can be written as
\begin{equation}
    \vect{d}(\vect{q}_g)\cdot \vect{d}(\vect{q}'_g)=\left(\vect{d}^0_g+\frac{1}{\sqrt{2}} \vect{d}^1_g \vect{u} \right)^2-\frac{1}{2}\left(\vect{d}_g^1 \vect{v}\right)^2.
\end{equation}
Finally, solving the remaining Gaussian integrals provides an explicit form for the time correlation function $C_{ge}$ as
\begin{equation}
    C_{ge}(\tau_g,\tau_e)= \left[\left(\vect{d}_g^0-\sum_{kp}\vect{d}^1_{g,k} C_{kp}^{-1}\Gamma_p\right)^2
         +\frac{\ii\hbar}{2}\sum_{kp}\vect{d}^1_{g,k} \cdot \vect{d}^1_{g,p}(C_{kp}^{-1}-D_{kp}^{-1}) \right] \chi(\tau_g,\tau_e) ,
         \label{eq:correlfinal}
\end{equation}
where the $n\times n$ matrices $\vect{C}$ and $\vect{D}$ and the vector $\bm{\Gamma}$ are respectively defined as
\begin{align}
&   \vect{C}=\vect{B}-\vect{A}, \\
& \vect{D} = \vect{B} + \vect{A},\\
& \bm{\Gamma} = \vect{J}^T \vect{E} \vect{K},  
\end{align}
and where
\begin{equation}
    \chi(\tau_g,\tau_e)=\sqrt{\frac{\det \vect{a}_g \det \vect{a}_e}{\det \vect{C} \det \vect{D}}}e^{\frac{\ii}{\hbar}\left(\vect{K}^T\vect{E}\vect{K}-\bm{\Gamma}^T\vect{C}^{-1}\bm{\Gamma}\right)} e^{-\ii \omega_{ge}\tau_e}.
\end{equation}
Note that the matrices $\vect{A}$, $\vect{B}$ and $\vect{E}$, hence $\vect{C}$, $\vect{D}$ and $\bm{\Gamma}$ as well, all depend on the parameters $\tau_g$ and $\tau_e$ through the definitions of $\vect{a}_g$, $\vect{a}_e$, $\vect{b}_g$ and $\vect{b}_e$ [Eqs.~(\ref{eq:amat}) and (\ref{eq:bmat})].
\section{Comparison between the computed and experimental absorption spectra of the D$_0$ $\rightarrow$ D$_2$ band of Np$^+$}\label{sec:app_absorption}

\begin{figure}
\includegraphics{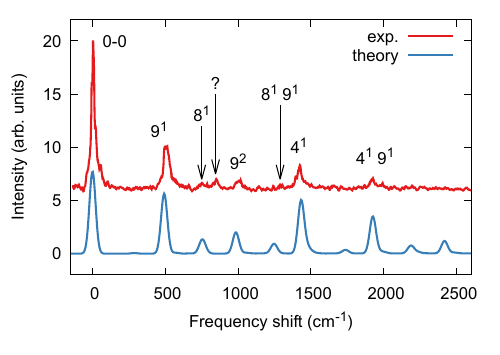}
\caption{Theoretical absorption spectrum of the D$_0$ $\rightarrow$ D$_2$  transition for Np$^+$ compared with the experimental photodissociation spectrum of  Np$^+$-Ar complex taken from  Ref.~\citenum{Pino:1999bh}}
\label{fig:abspino}
\end{figure}

Fig.~\ref{fig:abspino} shows a direct comparison between the theoretical absorption spectrum of the D$_0$ $\rightarrow$ D$_2$ band of Np$^+$ with the experimental photodissociation spectrum of the Np$^+$-Ar complex reproduced from Ref.~\citenum{Pino:1999bh}. To allow for a better comparison, both the theoretical and experimental spectra were shifted to have common origin bands, a Gaussian linewidth of 50~\cm\ and a uniform frequency scaling factor of $0.97$ were used in the calculation. The temperature was set to a value of $T=30$~K relevant to this experiment, although setting this parameter to 0 barely alters the computed spectrum.

In Fig.~\ref{fig:abspino}, we also report the assignment given in Ref.~\citenum{Pino:1999bh}, using the same vibrational frequency numbering. The positions of the various vibronic bands are well reproduced by our theoretical calculations. In particular, the experimental bands $9^1$, $8^1$, $9^2$, $8^1 9^1$, $4^1$, and $4^1 9^1$, located at $500$, $750$, $1001$, $1288$, $1417$, and $1941$~\cm, respectively, are found in the theoretical calculation at $492$, $758$, $984$, $1250$, $1432$, and $1925$~\cm. Note that the experimental vibronic band located at $845$~\cm, highlighted in Fig.~\ref{fig:abspino} by a question mark and whose assignment is unclear, is not observed in our calculation.

Residual differences in the intensities of the various vibronic bands are found between the experimental and theoretical spectra. However, the experimental spectrum does not result directly from an absorption process but rather from the photodissociation yield of van der Waals complexes. Such a process can lead to differences in the spectral intensity.

%
%
\section{Validity of the harmonic and HT approximations}
\label{sec:harmonic}

\begin{figure*}
    \centering
    \includegraphics[width=16.5cm]{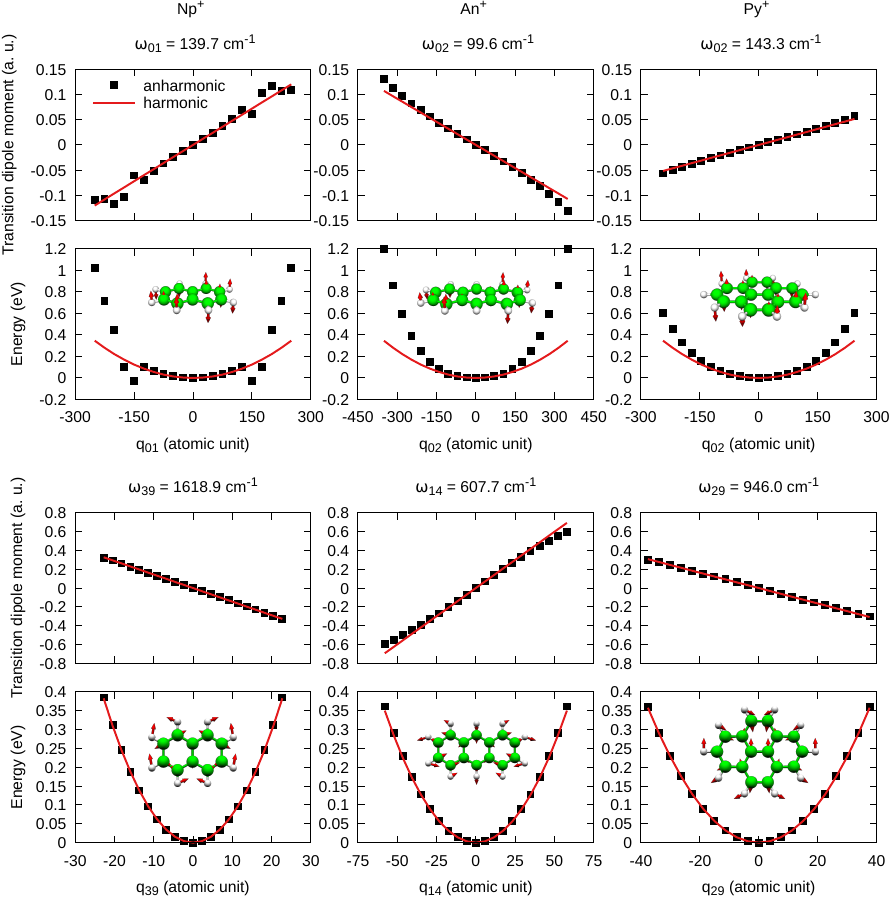}
    \caption{Comparison between TD-DFT calculations and the harmonic approximation for the energy of the D$_1$ electronic state and the linear approximation for the transition dipole moment along various normal mode coordinates of the naphthalene (left panels), anthracene (middle panels), and pyrene (right panels) cations. For each mode, atomic displacements are depicted as insets.}
    \label{fig:dipnrj}
\end{figure*}

The model for RF relaxation relies on the potential energy and transition dipole moment surfaces being correctly described by harmonic and linear expressions, respectively. Here we examine the quality of these approximations on selected electronic states of the three PAH cations.
Starting from the D$_1$ electronic excited state minimum geometry, a deformation was applied by following a specific eigenvector $q_{e,k}$ associated with mode $k$.
Because the extent of molecular deformation depends on the amount of energy injected in the molecule, a maximum deformation amplitude $q_{e,k}^{\rm max}$ was imposed as $2 \sqrt{\langle q_{e,k}^2 \rangle}$, where the canonical average $\langle \cdot \rangle$ is taken at a physically relevant temperature $T$ taken as 2000 K. In the harmonic approximation, this average depends on the mode frequency $\omega_{e,k}$ through
\begin{equation}
\left\langle q_{e,k}^2 \right\rangle = \frac{\hbar}{2\omega_{e,k}}\coth\left(\frac{\hbar \omega_{e,k}}{2 \boltz T}\right).
\end{equation}
Fig.~\ref{fig:dipnrj} shows the variations of the electronic excited state energy (relative to the minimum) and the transition dipole moment between the electronic ground and excited states, as a function of the normal mode coordinates $q_{e,k}$ and for selected modes of the three PAH cations, notably the (in-plane) modes having the highest contribution to the HT oscillator strength, as well as the lowest-frequency A$_u$ (out-of-plane) vibrational modes. In each graph, the values obtained from the TD-DFT calculations are compared with the predictions of the harmonic (energy) or linear (transition dipole moment) approximations for the corresponding quantities. For the transition dipole moment, which is a vectorial quantity, only the component that acquires a nonvanishing value after applying the deformation was considered. This component depends on the cation itself and on the symmetry of the normal mode.\par 
In the case of Np$^+$, and except for the softest A$_u$ mode $q_{e,1}$, the agreement between the direct calculation and the model system is very good, with almost indistinguishable differences.
In the specific case of $q_{e,1}$ for Np$^+$, and for large deformation amplitudes, the D$_0$ and D$_1$ electronic states cross each other. As a result, for large deformations $|q_{e,1}|\gtrsim 150$ a.u. the electronic ground state becomes the state with symmetry $^2$B$_{3u}$, while the electronic excited state switches to $^2$A$_u$ symmetry. This crossing at the origin of a conical intersection between the D$_1$ and D$_0$ states in Np$^+$ has been discussed earlier.\cite{Tokmachev:2008aa,Boggio-Pasqua:2019aa} Alongside the other conical intersections between the various electronic states of Np$^+$ and other PAHs, these features justify our statistical model in the microcanonical ensemble, energy redistribution between all electronic states occuring on fast time scales.
In the case of An$^+$ and Py$^+$, a similarly good agreement is found between the harmonic model and the TD-DFT calculations, with the exception of the lowest-frequency $A_u$ vibrational modes. In these cases, and while no crossing occurs between electronic states, very large intramode anharmonicities are found. Note that, as shown in Fig.~\ref{fig:modes}, all these low-frequency A$_u$ vibrational modes contribute only slightly to the overall HT oscillator strength.

\begin{acknowledgments}
Financial support by the ANR "INTERSTELLAR" under Grant No.
ANR-23-CE30-0023 is acknowledged.
The authors thank T. Pino for providing the experimental spectrum used in Fig. 11 and
acknowledge M. H. Stockett, H. Zettergren and J. N. Bull for useful discussions.  This work was performed using computational resources from the ``Mésocentre" computing center of Université Paris-Saclay, CentraleSupélec and École Normale Supérieure Paris-Saclay supported by CNRS and Région Île-de-France (https://mesocentre.universite-paris-saclay.fr/). 
\end{acknowledgments}

\section*{Author Declaration Section}

\subsection*{Conflict of Interest Statement }
The authors have no conflicts to disclose.

\subsection*{Author Contributions}
\noindent\textbf{Damien Borja:} Conceptualization (equal); Formal Analysis (lead); Investigation (equal); Methodology (equal); Software (equal); Visualization (lead); Writing -- Original Draft Preparation (equal); Writing -- Review \& Editing (equal)
\textbf{Florent Calvo:} Conceptualization (equal); Methodology (supporting); Writing -- Review \& Editing (equal)
\textbf{Pascal Parneix:} Conceptualization (equal); Funding Acquisition (equal); Methodology (supporting); Writing -- Review \& Editing (equal)
\textbf{Cyril Falvo:} Conceptualization (equal); Funding Acquisition (equal); Investigation (equal); Methodology (equal); Software (equal); Supervision (lead);  Writing -- Original Draft Preparation (equal);  Writing -- Review \& Editing (equal)
\section*{Data Availability}
The data and computational tools that support the findings of this study are available from the corresponding author upon reasonable request.

\bibliography{biblio}

\end{document}